\documentclass[aps,pra,twocolumn,superscriptaddress,showkeys,amsmath,amssymb,longbibliography]{revtex4-1}		
\usepackage{graphicx}
\usepackage{dcolumn}
\usepackage{bm}
\usepackage[colorlinks,urlcolor=blue,citecolor=blue,linkcolor=blue]{hyperref}
\usepackage{subfig}
\usepackage[font=small,labelfont=bf,format=plain,justification=centerlast,labelsep=period]{caption}
\usepackage{tikz}
\usetikzlibrary{arrows,shapes,positioning,shadows,svg.path,backgrounds,fit}

\usepackage{verbatim}
\usepackage[normalem]{ulem}
\usepackage{orcidlink}

\begin{document}



\title{Enhanced steady-state coherence via repeated system-bath interactions}

\author{Ricardo Rom\'an-Ancheyta\,\orcidlink{0000-0001-6718-8587}}
\email{ancheyta6@gmail.com}
\address{Instituto Nacional de Astrof\'isica, \'Optica y Electr\'onica, Calle Luis Enrique Erro 1, Sta. Ma. \\ Tonantzintla, Puebla CP 72840, Mexico}
\author{Michal Kol\'a\ifmmode \check{r}\else \v{r}\fi{}\,\orcidlink{0000-0003-0860-3212}}
\email{kolar@optics.upol.cz} 
\affiliation{Department of Optics, Palack\'y University, 17. listopadu 1192/12, 771 46 Olomouc, Czech Republic}
\author{Giacomo Guarnieri\,\orcidlink{0000-0002-4270-3738}}
\email{gguarnieri88@gmail.com}
\affiliation{School of Physics, Trinity College Dublin, College Green, Dublin 2, Ireland}
\author{Radim Filip\,\orcidlink{0000-0003-4114-6068}}
\email{filip@optics.upol.cz}
\affiliation{Department of Optics, Palack\'y University, 17. listopadu 1192/12, 771 46 Olomouc, Czech Republic}
\date{\today}

\begin{abstract}
The appearance of steady-state coherence (SSC) from  system-bath interaction proves that quantum effects can appear without an external drive. Such SSC could become a resource to demonstrate quantum advantage in the applications. 
We predict the generation of SSC if the target system {\em repeatedly} interacts with independent and non-correlated bath elements. 
To describe their behavior, we use the collision model approach of system-bath interaction, where the system interacts with one bath element (initially in an incoherent state) at a time, asymptotically (in the fast-collision regime) mimicking a macroscopic Markovian bath coupled to the target system. Therefore, the SSC qualitatively appears to be the same as if the continuous Markovian bath would be used. We confirm that the presence of \textit{composite} system-bath interactions under the rotating-wave approximation (RWA) is the necessary condition for the generation of SSC using thermal resources in collision models. 
Remarkably, we show that SSC substantially increases if the target system interacts {\em collectively} with more than one bath element at a time. Already few bath elements collectively interacting with the target system are sufficient to increase SSC at non-zero temperatures at the cost of tolerable lowering the final state purity. From the thermodynamic perspective, the SSC generation in our collision models is inevitably linked to a non zero power input (and thus heat dissipated to the bath) necessary to reach the steady-state, although such energetic cost can be lower compared to cases relying on SSC non generating interactions. 
\end{abstract}

\maketitle

\section{Introduction}
\label{intro}

It is well known that quantum coherence is a 
valuable physical resource useful for many 
applications~\cite{AdessoRMP2017}.
In quantum thermodynamics, for example, 
experiments have demonstrated~\cite{UzdinPRL2015} that, 
within the small-action limit~\cite{Uzdin2015}, quantum 
coherence between different internal energy states of 
the working substance allows a quantum heat engine to 
produce more power than its classical counterpart.
In quantum metrology, it has been shown~\cite{Smirne_2019} 
that long-time coherence in the state of the sensing 
particles can be used to outperform  the precision of 
frequency estimation~\cite{Smirne_2016} when compared with entanglement-based strategies. However, such strategy relies on the coherence trapping effect~\cite{SabrinaPRA2014} and, 
therefore, has the practical disadvantage that the state of
the probes needs some initial coherence. Therefore, such quantum advantage cannot appear autonomously in quantum matter. 

It is precisely the aim of several investigations to find 
processes in microscopic and mesoscopic systems that lead, on demand and without external coherent drives,
to the generation of robust quantum coherence, entanglement in the steady-state or quantum synchronization \cite{PhysRevA.100.012133}. 
For instance, in~\cite{Bohr_Brask_2015} an autonomous quantum 
thermal machine produces {\em degenerate} steady-state 
coherence (SSC) in a two-qubit system interacting, 
incoherently, with two thermal baths at different temperatures.
In~\cite{SSC_Radim}, sufficient conditions for the generation 
of {\em energetic} SSC (coherence between states with 
different energies~\cite{ManzanoPRE19}) in a two-level 
system in contact with a single thermal bath were identified. 
Those conditions that we will discuss in detail in the present work rely on the particular structure of the composite~\cite{SSC_Radim} system-bath interaction.
Remarkably, in both examples~\cite{Bohr_Brask_2015,SSC_Radim},
the SSC are independent of the initial state of the system, which could be initially incoherent.
\textcolor{black}{Such independence makes these strategies, in a sense, similar to earlier proposals for preparing non-equilibrium quantum phases~\cite{Diehl2008} and implementing robust dissipative quantum computation~\cite{Verstraete2009} using quantum-reservoir engineering of many-body systems.}

The framework put forward in~\cite{SSC_Radim} was recently
applied in~\cite{guarnieri2020non} to obtain non-equilibrium steady-states (NESS) with SSC. There, the thermodynamic cost to produce such coherence was calculated numerically and, interestingly, 
non-zero work and heat currents at the steady-state were necessary to maintain the NESS with SSC~\cite{guarnieri2020non}.
On the other hand, in~\cite{Purkayastha2020}, an experimentally feasible semiconductor double-quantum dot charge qubit in permanent contact with a thermal bath was proposed to implement the characteristic structure of the interaction Hamiltonian of~\cite{SSC_Radim}.

Other works deal with the characterization of coherence but from the point of view of the resource theory of quantum thermodynamics~\cite{RPP_Lostaglio_2019,Horodecki2013,Fernando_PRL_2013,Fernando_PNAS_2015,PRX_Muller_2018}. For instance, it was shown in~\cite{NJP_Faist_2015} that Gibbs-preserving maps can outperform thermal operations by creating quantum coherence from energy eigenstates. Remarkably, in~\cite{PRL_Muller_2019} a crucial no-go theorem was introduced, showing that quantum coherence cannot be broadcast in every-finite dimensional system, therefore, ruling out the free cost generation of coherent superpositions from incoherent states; see also~\cite{PRL_Iman_2019} for a closely related work.

In this paper, motivated by the generality of the sufficient 
conditions that guarantee the generation of SSC 
in~\cite{SSC_Radim}, {\color{black}{\it we extend those results} }to the framework 
of repeated, pulsed interactions~\cite{Barra2015,Esposito_PRX_17,Pereira,StrasbergPRL2019,Stefan19}, also known as
collision models~\cite{ciccarello2021quantum,Ciccarello13,Ciccarello17,ciccarello2017collision,rodrigues2019,Onur2020,Heineken}. 
These models not only give theoretical insight to microscopic processes in the baths required to achieve SSC, but mainly they can be\textcolor{black}{, in some cases, efficiently implemented on a quantum processor~\cite{Maniscalco_npj_20,Cattaneo_PRL_21} or linear optical schemes~\cite{Cuevas_SR_2019}; other potential platforms may be cold trapped ions~\cite{FluhmannNature2019} and quantum circuits~\cite{campagneibarcq2019quantum}.}
Such proof-of-principle {\color{black}non-autonomous} experiments will simulate SSC with current experimental techniques and verify {\color{black}mechanisms to obtain} of SSC under various conditions in parallel with the ongoing search for suitable autonomous platforms \cite{Purkayastha2020}.

In particular, we study here the creation and {\em collective} 
enhancement of energetic SSC (along with high purity) in a \textcolor{black}{target system (described by a qubit or by a harmonic oscillator)} interacting with an effective bath. The effective bath is modeled as a stream of bath elements, clusters of qubits or linear harmonic oscillators in thermal states that interact for a short period of time with the target system in which SSC is to be created.
This underlying microscopic procedure yields effectively a Markovian time-independent master equation description of the target system. An analytic solution is employed to access the steady-state\textcolor{black}{, independent of the system's initial state.} In order to obtain the transient dynamics we use numerical calculations, as well as approximated solutions. 

\subsection*{Overview of the results}
Our results show that even for the composite system-bath interactions taken into account, there exist \textcolor{black}{specific} scenarios \textcolor{black}{(for instance, when the RWA is not valid)} in which the sufficient conditions found in~\cite{SSC_Radim} are not applicable for SSC generation in the context of collision models describing open quantum dynamics. This is due to the physically different way of modelling the thermal bath in this work, compared to \cite{SSC_Radim}, showing that these two approaches are not equivalent from the perspective of SSC generation.

We find that at low bath  temperatures

\textcolor{black}{$i)$} \textcolor{black}{Energetic SSC and the corresponding purity} of the system \textcolor{black}{reach their} maximum value. 

\textcolor{black}{$ii)$} \textcolor{black}{Energetic SSC} can be substantially increased, together with the system energy, as the number of elements in the corresponding bath-clusters increases, although lowering the resulting system purity at the same time. 

\textcolor{black}{$iii)$ Both energetic SSC and purity possess a small constant plateau, allowing for possible experimental observation.} 

For high bath temperatures~\cite{Latune_2019}

\textcolor{black}{$iv)$} The SSC is washed out and the target system reaches a completely incoherent mixed state, irrespective of the bath-cluster size.

To characterize our models from a thermodynamic perspective, we establish a clear connection between the power input, characterizing the steady state, and created non-zero SSC for certain class of collision models used in our work. Our results show that for, e.g., qubit-qubit collisions defined by certain interaction, the power input is proportional to the created SSC. On the other hand, other class of interactions, not generating SSC, needs as well a positive power supply. These facts lead us to an observation that 

\textcolor{black}{$v)$} Positive power input does not guarantee SSC  generation, but any SSC is consuming certain power to be generated.

\textcolor{black}{Finally, we find that the correct splitting, in terms of heat and work, in the dynamical version of the First Law of thermodynamics cannot be obtained by just knowing the change of the system's internal energy. Besides, the generation of energetic SSC induces substantial modifications in the heat current that deviate from the Landauer formulation of transport theory.}
\section{SSC FROM QUBIT BATH ELEMENTS}
\label{TLS}
\subsection{Asymptotic Coherence}
\label{TLS-steady}
\begin{figure}[t]
\includegraphics[width=.95\columnwidth]{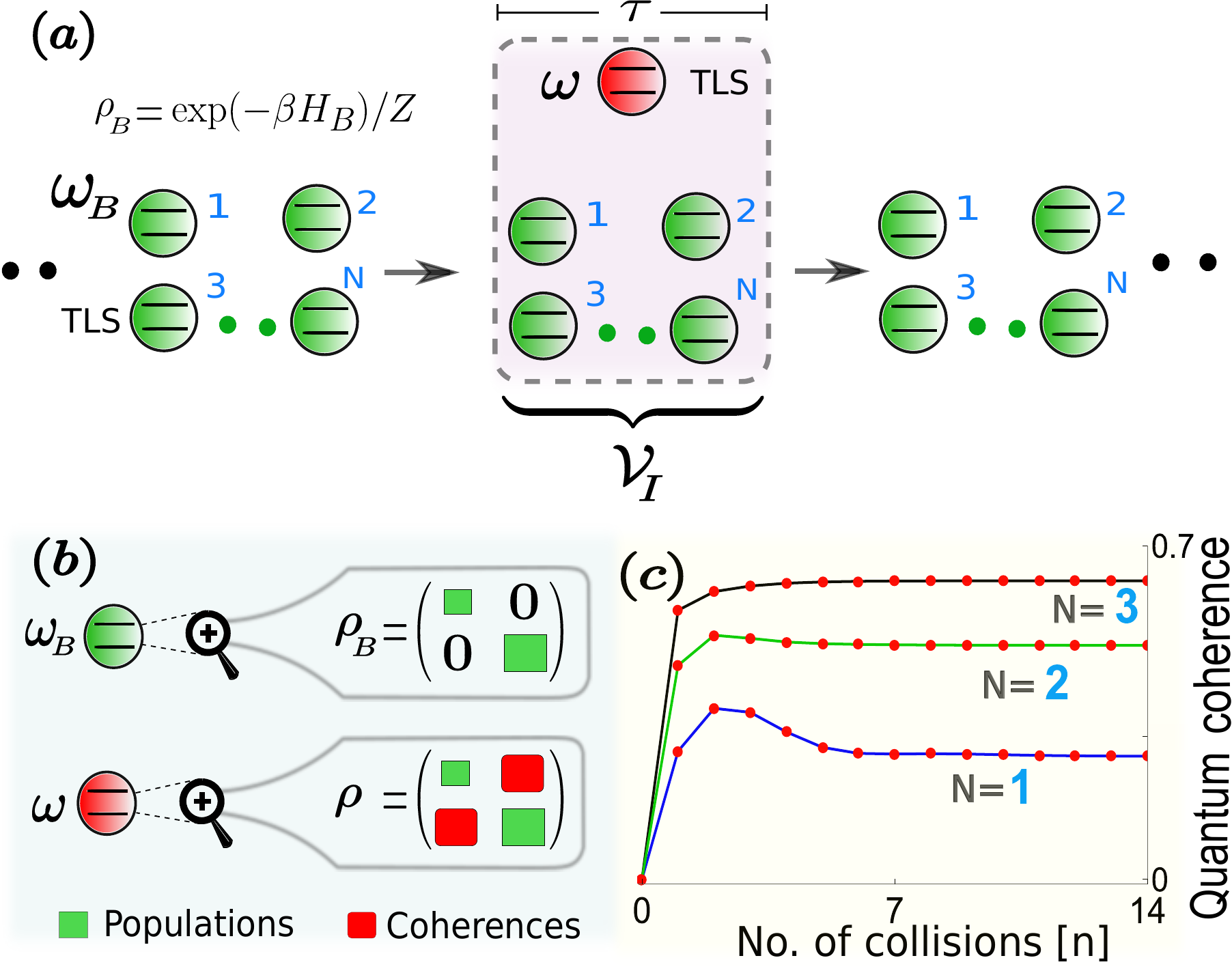} 
\caption{
Schematic representation of a basic collision model or repeated interaction scheme extended  throughout the paper. {\bf (a)} The bath elements represented by clusters of $N$ independent and non-correlated two-level systems (TLS) or qubits (lower green circles) of frequency $\omega_B$ are initially in a thermal state of inverse temperature $\beta$. Each bath element interacts collectively during a short period of time, $\tau$, with a target TLS (red) of the frequency $\omega$ through the interaction Hamiltonian $\mathcal{V}_I$. This procedure makes short-time collisions with the bath elements acting as an effective Markovian bath \cite{guarnieri2020non}, although such collision-schemes have the potential to generate more general types of evolution. {\bf (b)} Pictorial representation of the initial density matrix $\rho_B$ of each bath TLS. Before the interaction with the target qubit, $\rho_B$ has only diagonal elements (green squares). After a few  collisions, the density matrix $\rho$ of the target TLS will contain coherence (red squares) regardless if $\rho$ was, initially, an incoherent state. {\bf (c)} Due to the composite structure of $\mathcal{V}_I$, the coherence generated in $\rho$ reaches, after several collisions, a stationary value. Remarkably, such steady-state coherence (SSC) substantially increases with the size of each cluster. It is shown how fast typically SSC, quantified with the $l_1$-norm of coherence, grows for $N=\{1,2,3\}$, $\tau=1$ and $\beta=5$.
}
\label{setup_clusters}
\end{figure}

\begin{figure*}
\subfloat[\label{normCoheQuQu}]{
\begin{tikzpicture} 
  \node (img1) {\includegraphics[width=.93\columnwidth]{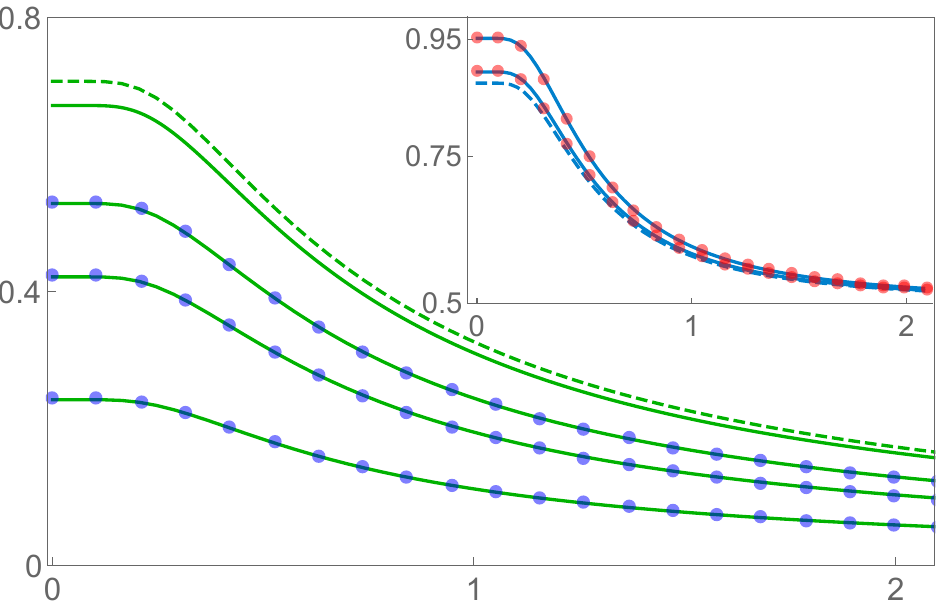}};
  \node[above=of img1, node distance=0cm, yshift=-6.8cm,xshift=0cm] {$k_BT/\hbar\omega_B$};
  \node[above=of img1, node distance=0cm, yshift=-3.3cm,xshift=-3.1cm] {$N=3$};
  \node[above=of img1, node distance=0cm, yshift=-2.5cm,xshift=-3.1cm] {$N=8$};
  \node[above=of img1, node distance=0cm, yshift=-3.5cm,xshift=.8cm] {$N\gg 1$};
  \node[above=of img1, node distance=0cm, yshift=-1.8cm,xshift=-3.1cm] {$N\gg 1$};
  \node[above=of img1, node distance=0cm, yshift=-4.1cm,xshift=-3.1cm] {$N=2$};
  \node[above=of img1, node distance=0cm, yshift=-5.2cm,xshift=-3.1cm] {$N=1$};
   \node[above=of img1, node distance=0cm, yshift=-2.cm,xshift=1.3cm] {$N=1$};
   \node[above=of img1, node distance=0cm, yshift=-2.1cm,xshift=-1.3cm] {RWA interaction};
  \node[above=of img1, node distance=0cm, yshift=-1.2cm,xshift=.2cm] {$\mathcal{P}_{\rm ss}$};
   \node[above=of img1, node distance=0cm, yshift=-1.2cm,xshift=-3.1cm] {{\color{black}$\mathcal{C}_{\rm ss}$}};
\end{tikzpicture}}
\subfloat[\label{fig-TLS-cohmax}]{
\begin{tikzpicture} 
  \node (img1)  {\includegraphics[width=.93\columnwidth]{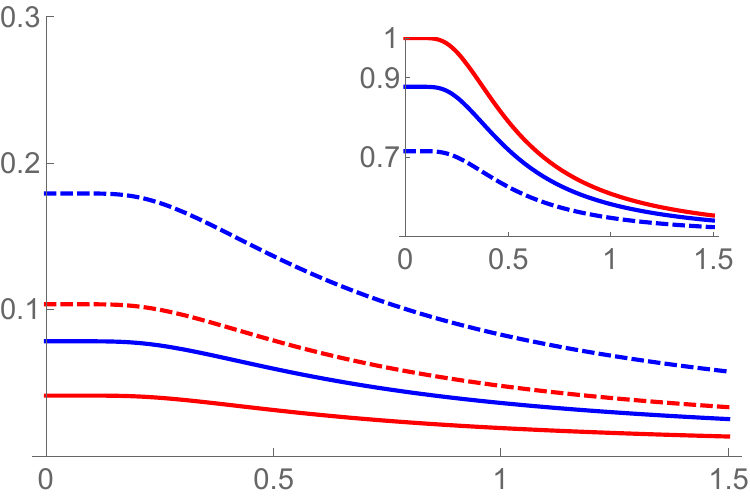}};
  \node[above=of img1, node distance=0cm, yshift=-6.8cm,xshift=0cm] {$k_BT/\hbar\omega_B$};
  \node[above=of img1, node distance=0cm, yshift=-3.2cm,xshift=-2.7cm] {$N_{\rm C-R}=3$};
  \node[above=of img1, node distance=0cm,rotate=-30,anchor=center, yshift=-2.2cm,xshift=2.5cm] {$N=3$};
  \node[above=of img1, node distance=0cm, yshift=-4.3cm,xshift=-2.7cm] {$N_{\rm RWA}=3$};
  \node[above=of img1, node distance=0cm, yshift=-5.3cm,xshift=-2.7cm] {$N_{\rm C-R}=1$};
  \node[above=of img1, node distance=0cm,rotate=-30,anchor=center, yshift=-1.7cm,xshift=1.9cm] {$N=1$};
  \node[above=of img1, node distance=0cm, yshift=-5.8cm,xshift=-2.7cm] {$N_{\rm RWA}=1$};
   \node[above=of img1, node distance=0cm, yshift=-2.cm,xshift=1.9cm] {$N=\{1,3\}$};
  \node[above=of img1, node distance=0cm, yshift=-1.5cm,xshift=.8cm] {$\overline{\mathcal{P}}_{\rm TS}^{\rm TLS}$};
  \node[above=of img1, node distance=0cm, yshift=-3.7cm,xshift=.8cm] {{\color{blue}C-R}};
  \node[above=of img1, node distance=0cm, yshift=-2.5cm,xshift=2.25cm] {{\color{red}RWA}};
   \node[above=of img1, node distance=0cm, yshift=-1.5cm,xshift=-3.cm] {{\color{black}${\overline{\mathcal{C}}}_{\rm TS}^{\rm TLS}$}};
\end{tikzpicture}}
\caption{{\bf (a)} Steady-state coherence (SSC, Eq.~(\ref{cohe_ss}))
in the target two-level system of Fig.~\ref{setup_clusters} 
as a function of scaled temperature of the bath elements (lowest green solid line) in the case of RWA type of interaction, Eq.~\eqref{ham:int:qubit:cluster} for $N=1$. When the target qubit interacts collectively with clusters of $N$ non-correlated bath qubits (see Fig.~\ref{setup_clusters}), the SSC increases substantially~(upper green solid lines) until it saturates to the value~$\mathcal{C}_{\rm ss}$ (green dashed line). The latter corresponds to the theoretical upper 
bound~$\mathcal{C}_{\rm ss}=\mathcal{C}_0\tanh\big(\omega_B/(2k_BT)\big)$ 
maximizing SSC for $N\gg 1$. Parameters are: $\omega=\omega_B=1$, 
$f_1=f_2/\sqrt{2}$, $f_2=0.6$ and $N=\{1,2,3,8\}$. Inset: 
steady-state purity of the target qubit as a function 
of the same scaled temperature. Purity decreases when the 
number of qubits in the clusters increases from $N=\{1,2\}$ 
(blue solid lines) until a saturated value 
(blue dashed line) for $N\gg 1$. 
Blue and red dots represent a purely numerical calculation of the repeated interaction model where, for $\tau=0.051$, the steady-state is reached after~$\sim 10^3$ collisions. 
Note that these results are independent of the initial 
state of the target qubit and that the $\mathcal{C}_{\rm ss} $ and $\mathcal{P}_{\rm ss}$ have opposite trends of cluster-size $N$ dependence, i.e., as the system coherence increases with the size of the cluster, its purity decreases. Remarkably, the plateau region in $\mathcal{C}_{\rm ss}$ and $\mathcal{P}_{\rm ss}$ allows reaching their maximum values for $k_BT/\hbar\omega_B>0$. {\bf (b)} The dependence of optimized transient state coherence (TSC) $\overline{\mathcal{C}}_{\rm TS}$, Eq.~\eqref{eq-trans-coh-approx-TLS}, on the bath temperature in cases when the system interacts with $N=\{1,3\}$ bath TLS via RWA interaction (labeled by superscript RWA), Eq.~\eqref{ham:int:qubit:cluster}, or with counter-rotating (C-R) terms included, Eq.~\eqref{hami_int_qubit_rabi_type}. The (C-R) results clearly have an edge over the RWA results in terms of attainable coherence $\overline{\mathcal{C}}_{\rm TS}$. On contrary, the corresponding optimized system state purity $\overline{\mathcal{P}}_{\rm TS}$, Eqs.~\eqref{eq-trans-purity-TLS}, of (C-R) interaction is suppressed with respect to the RWA scenario. The temperature dependence is entering the results through the system initial inversion $z_0=-\tanh(\hbar\omega/(2k_BT))$ and assumption that the system and the bath have initially the same temperature $T$ and frequency $\omega=\omega_B$. The values of the parameters are $\omega_B=\omega=1$, $f_1=f_2=0.15$. These values of the interaction constants (while being close to the edge of the validity of approximation \eqref{eq-trans-coh-approx-LHO} ) are roughly three times smaller than the values optimizing $\mathcal{C}_{\rm SS}$ in panel {\bf (a)}. This is the reason for lower values reached in the transient regime.}
\end{figure*}
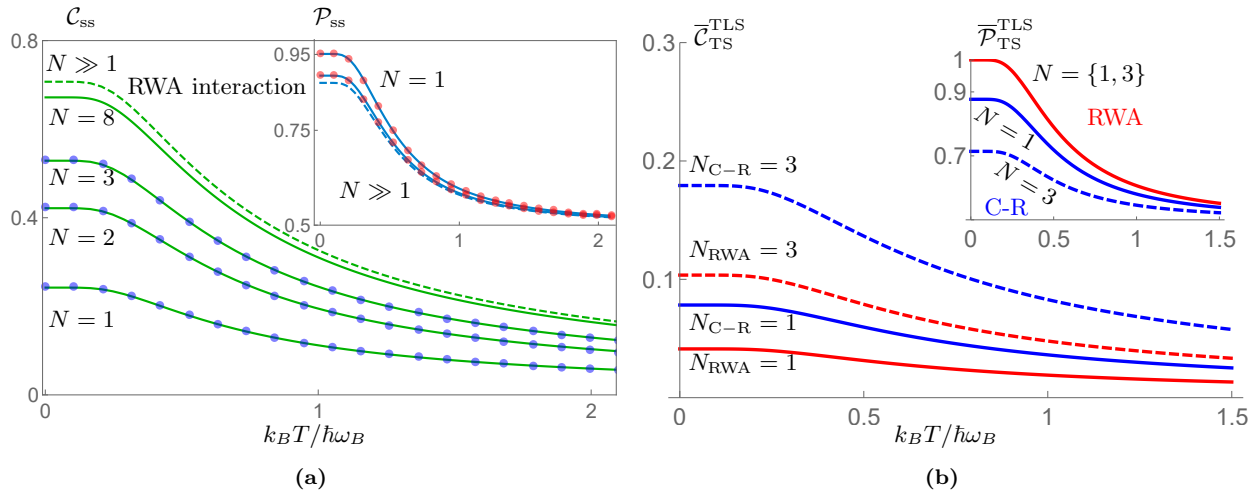

In this section we focus on the setup described in Fig.~\ref{setup_clusters} with individual interactions described by $N=1$, i.e., when the target system consists of a two-level system (TLS, qubit) with a free Hamiltonian $\mathcal{H}_S=\omega\sigma_z/2$ and the bath elements are represented by other TLS of frequency $\omega_B$. Here and from now on, we set $\hbar=1$. We assume that the interaction between the target qubit and each respective bath element can be written in the particular form, which we call {\it composite}, allowing for spontaneous creation of steady state coherence (SSC)~\cite{SSC_Radim}:  
\begin{equation}\label{hami_int_qubit_qubit}
\mathcal{V}_I=f_1\sigma_z\otimes(\sigma_-^B+\sigma_+^B)
+f_2(\sigma_+\otimes\sigma_-^B+\sigma_-\otimes\sigma_+^B),
\end{equation}	
where $\sigma_{\pm}$ ($\sigma_{\pm}^B$) are the ladder operators of the  target TLS (bath element) and $f_{1,2}$ are two real coupling constants. 
%
In the above expression, the composite interaction  $\mathcal{V}_I$ consists of the so called parallel, 
$\mathcal{H}^{\parallel\mathcal{H}_S}\textcolor{black}{=f_1\sigma_z\otimes(\sigma_-^B+\sigma_+^B)}$, and orthogonal, $\mathcal{H}^{\perp \mathcal{H}_{S}}\equiv\mathcal{V}_I-\mathcal{H}^{\parallel\mathcal{H}_S}$, 
components with respect to $\mathcal{H}_{S}$~\cite{SSC_Radim}, respectively. The parallel component alone ($f_1\neq 0$, $f_2=0$) induces dephasing of the target qubit while creating coherence on the bath element. On the other hand, the orthogonal component alone ($f_1=0$, $f_2\neq 0$) describes a damping
interaction between the target system and the bath element,
i.e., it causes solely quanta hopping, while the total excitation being conserved. 
Such hopping alone (the orthogonal part of \eqref{hami_int_qubit_qubit}) can create coherence only between the incoherent target qubit and the incoherent bath element. Only if the bath element has quantum coherence on its own, it can be transferred to the target by the hopping interaction. In order to have non-zero quantum coherence in the target
asymptotic steady-state, both interactions are therefore necessary in the interaction Hamiltonian; it is however not clear if they are also sufficient for SSC creation.
Only if both components are turned on ($f_1\cdot f_2\neq 0$), the coherence in the bath element (created by the parallel component) is transferred to the target system by the orthogonal interaction. The interaction \eqref{hami_int_qubit_qubit} approximately describes pulsed dynamics of the two-level system in trapped ion \cite{FluhmannNature2019} and superconducting circuit \cite{campagneibarcq2019quantum} experiments when the oscillator representing bath $B$ is weakly excited. It corresponds to the low-temperature limit, where SSC appears. At this point we would like to mention the general logic adopted throughout this work. Each collision defined by the interaction $\mathcal{V}_I$ corresponds, in principle, to a certain microscopic model, underlying some type of experiment (e.g., \cite{FluhmannNature2019,campagneibarcq2019quantum}). We model such discrete type of dynamics numerically, employing short-time (but finite) interactions (collisions) of the target system with a bath unit. Every such particular target-bath interaction $\mathcal{V}_I$ defines, under certain conditions, corresponding Markovian master equation (ME), see Appendix~\ref{model} for derivation, which is more suitable for analytical solutions used and discussed in our work. We have checked the correspondence of the purely numerical and ME-based (analytical) results, confirming excellent match in the regime of parameters used throughout the paper. Examples of this match are shown, e.g., in Figs.~\ref{normCoheQuQu}~or~\ref{bloch_evolution} as discrete dots on the top of continuous curves. 

Notice that Eq.~(\ref{hami_int_qubit_qubit}) can be rewritten as a bi-linear combination $\mathcal{V}_I=s^\dagger\otimes A+s\otimes A^\dagger$ between system and bath operators, if we define the operators as $s=f_1\sigma_z+f_2\sigma_-$ and $A=\sigma_-^B$. 
If we perform the corresponding trace over the incoherent bath states $\rho_B$,   
the dynamical equation for the target qubit acquires the well-known form of the following time-independent Markovian master equation (see Appendix~\ref{model} for a detailed derivation):
\begin{equation}
\begin{split}
\frac{{ d}\rho}{{ d}t}=-\frac{i\omega}{2}[\sigma_z,\rho]+\langle \sigma_-^B\sigma_+^B\rangle\mathcal{L}[f_1\sigma_z+f_2\sigma_-]\rho\qquad \\ 
+ \langle \sigma_+^B\sigma_-^B \rangle\mathcal{L}[f_1\sigma_z+f_2\sigma_+]\rho,
\label{eq:mast:2}
\end{split}
\end{equation}
where $\mathcal{L}[x]\rho\equiv x\rho x^\dagger-\frac{1}{2}(x^\dagger x\rho+\rho x^\dagger x)$ is the usual Lindblad super-operator and $\langle x \rangle={\rm tr}\{x\rho_B\}$ is the expectation value of an arbitrary operator $x$ with respect to the initial (thermal) bath state. In addition to the detailed derivation of Eq.~\eqref{eq:mast:2} given in Appendix~\ref{model}, we stress here that this equation holds conditioned on the limit of short interaction time $\tau$ {\it and} the condition of re-normalization of $\mathcal{V}_I$ by $1/\sqrt{\tau}$, see Appendix \ref{model} for details. 
The above equation can be solved easily by numerical 
methods. Exactly both of the above mentioned conditions allows for direct connection of the collision model and system dynamics described by an effective master equation of the Lindblad type. This connection allows for the possibility to obtain analytical
expressions for $\langle \sigma_x\rangle$,
$\langle \sigma_y\rangle$ and $\langle \sigma_z\rangle$, in the steady state (see appedix~\ref{apx_bloch}). It is important to note that the second and third term in the right-hand side of~(\ref{eq:mast:2}) should not be interpreted as terms causing only incoherent de-excitation and incoherent excitation, respectively. As we will see, these Lindblad super-operators $\mathcal{L}[x]\rho$  are able to generate coherence in the energy basis of the target qubit, even in the steady state, because they contain a linear combination of both parallel ($f_1\sigma_z$) and orthogonal ($f_2\sigma_\pm$) components with respect to $\mathcal{H}_S$ as their argument $[x]$. 
We recall that for $f_1\cdot f_2=0$, Eq.~(\ref{eq:mast:2}) will {\it not} generate SSC. 

In order to quantify the possible generation of coherence in the target (qubit) system, we use the $l_1$-norm of coherence measure~\cite{PlenioCohe}. This is defined as the absolute value of the off-diagonal element of the density matrix of interest \textcolor{black}{$\mathcal{C}(t)={\sum}_{i\neq j}|\rho_{i,j}(t)|.$}
For the state of the qubit $\rho$ this can be easily written as \textcolor{black}{$\mathcal{C}(t)=|\langle \sigma_x(t)\rangle+i\langle \sigma_y(t)\rangle|$,}
used from now on, having the following form in the steady-state, taking $\mathcal{C}_{\rm ss}\equiv\lim_{t\rightarrow\infty}\mathcal{C}(t)$  (see appedix~\ref{apx_bloch_jcm} 
for details)
\begin{equation}\label{cohe_ss}
\mathcal{C}_{\rm ss}=f_1f_2\frac{r(T)}{s(T)+\omega^2},
\end{equation}
where
\begin{equation}\label{r_function}
r(T) =\langle [\sigma_-^B,\sigma_+^B]\rangle\big[\omega^2+\langle\{\sigma_-^B,\sigma_+^B\}\rangle^2 (2f_1^2+f_2^2/2)^2\big]^\frac{1}{2},  
\end{equation}
\begin{equation}\label{s_function}
s(T) =\langle\{\sigma_-^B,\sigma_+^B\}\rangle^2(2f_1^2+f_2^2/2)(f_1^2+f_2^2/2), 
\end{equation}
are two functions that may depend, based on the result of the commutator and anti-commutator average of the bath operators, on the temperature $T$ of the corresponding bath elements. In deriving Eq.~(\ref{eq:mast:2}), we use the condition ${\rm tr}_{\rm B}\{\mathcal{V}_I\rho_B\}=0$, which implies that the initial state of the bath elements can not have coherence in the energy basis. Moreover, any diagonal state of the ancillary qubits is a Gibbs thermal state since it can always be written in the Gibbs form $\rho_{\rm th}^B=\exp(-\beta\omega_B\sigma_z^B/2)Z^{-1}$,
where $Z=2\cosh(\beta\omega_B/2)$ and $\beta=(k_BT)^{-1}$ is the inverse temperature.
This temperature is the so called apparent temperature introduced in~\cite{Latune_2019} as $T\equiv(\omega_B/k_B)\ln(p_g^B/p_e^B)^{-1}$,
where $p_g^B$ ($p_e^B$) is the probability to find each bath qubit in its ground (excited) state. 
In such case $\langle[\sigma_-^B,\sigma_+^B]\rangle=\tanh(\beta\omega_B/2)$ and $\langle\{\sigma_-^B,\sigma_+^B\}\rangle=1$. This makes $r(T)$ the only temperature dependent function. In particular, when $\beta\ll 1$ we approximate $\tanh(\beta\omega_B/2)\approx\beta\omega_B/2$ and the steady state coherence, Eq.~(\ref{cohe_ss}), vanishes approximately as $\omega_B(\omega T)^{-1}$ in the high temperature limit. The opposite low-temperature limit, $\beta\gg 1$, leads the only thermal factor to $\tanh(\beta\omega_B/2)\approx 1$, leaving only the rest of the parameters to determine the SSC value.
From Eq.~(\ref{cohe_ss}) it is interesting to note that, as long as the product $f_1\cdot f_2$ is nonzero,  SSC can be generated in the target qubit, even at zero temperature, see green lower curve in Fig.~\ref{normCoheQuQu}. 
Ultimately, this result (\ref{cohe_ss}) is independent of whether we had chosen $\sigma_y^B$ instead of $\sigma_x^B$ as the parallel component $\mathcal{H}^{\parallel \mathcal{H}_S}$ of (\ref{hami_int_qubit_qubit}).

We would like to stress that the result of generation of SSC is completely independent of the initial state of the target system and it is very different from the results obtained in~\cite{Angsar17,ccakmak2020ergotropy,LatunePRA2019,LatunePRR2019}. In those works it is not possible to create SSC at zero temperature, because their strategy relies on the presence of thermal photons that must be absorbed by a composite (many-body) system made of, at least, two coupled two-level atoms (the target system) in which the SSC are to be created. Moreover, such atoms had to be close enough in space, in order to treat them as indistinguishable, when a thermal photon was absorbed.
This is in contrast with our repeated interaction scheme, where SSC can be generated on a single two-level system and even at zero temperature.

Interestingly, following the ideas of~\cite{Angsar19,Deniz19,roman2019spectral}, 
it is instructive to generalize previous 
results for the case in which the target qubit interacts, repeatedly, with {\em clusters} made of $N$ independent and non-correlated bath qubits \cite{Gangloff62}, instead of a single bath TLS, see Fig.~\ref{setup_clusters}.
Thus, in Eq.~(\ref{hami_int_qubit_qubit}) we can replace $\sigma_\pm^B$ by $ \sum_{j=1}^N\sigma_\pm^{(j)}\equiv S_\pm$, thus the corresponding collective interaction between the target qubit and each cluster reads
\begin{equation}\label{ham:int:qubit:cluster}
\mathcal{V}_I=f_1\sigma_z\otimes(S_-+S_+)
+f_2(\sigma_+\otimes S_-+\sigma_-\otimes S_+).
\end{equation}
The operators $S_\pm$ are known as the collective spin operators~\cite{Angsar19}.
\textcolor{black}{Although such a type of collective interaction could be challenging to implement experimentally in the context of collision models, recent works show that clusters with up to $N=25$ fully controllable superconducting qubits can be  realized~\cite{Ustinov_2021,Wang_PRL_2021}.}
For this new collective interaction the basic structure of 
(\ref{eq:mast:2}) 
and (\ref{cohe_ss}) 
will essentially remain unchanged.
In such case, it is easy to show that the expectation value of the commutator and anti-commutator between the collective spin operators, and with respect to the incoherent cluster state $\rho_{\rm cl}=\bigotimes_{j=1}^N\rho_{\rm th}^j$  ($\rho_{\rm th}^j$ being the thermal state of the $j$-th qubit in the cluster) is
\begin{equation}\label{conm_rel_spins}
\langle\{S_-,S_+\}\rangle=N,\quad
\langle[S_-,S_+]\rangle=N\tanh(\beta\omega_B/2).
\end{equation}
Replacing these expressions in (\ref{r_function}) 
and (\ref{s_function}), the $l_1$-norm of coherence~(\ref{cohe_ss}) will now also depend on $N$ as
\begin{equation}\label{cohe_ss_qq}
\mathcal{C}_{\rm ss}
=f_1f_2\tanh(\beta\omega_B/2)\frac{r(N)}{s(N)+\omega^2},
\end{equation}
where 
\begin{subequations}
\begin{align}
r(N) &=N\sqrt{\omega^2+N^2 (2f_1^2+f_2^2/2)^2},\label{r_Nfunction_N}\\
s(N) &=N^2(2f_1^2+f_2^2/2)(f_1^2+f_2^2/2).\label{s_function_N}
\end{align}
\end{subequations}

It is worth noting that in this cluster scenario, substantial increase of the SSC values in the target qubit can be obtained when the size of each cluster also increases. The behaviour of such SSC, as a function of the bath temperature and the number of bath qubits in each cluster is shown in Fig.~\ref{normCoheQuQu}. When the number of qubits in the clusters is large, $N\gg1$ (upper index ``$\infty$" in Eq.~\eqref{cohe_max}), the steady-state coherence (\ref{cohe_ss_qq}) can be well approximated by a simple form
\begin{equation}\label{cohe_max}
\mathcal{C}_{\rm ss}^{\infty}\approx\mathcal{C}_0\tanh(\beta\omega_B/2),\quad\mathcal{C}_0\equiv\frac{f_1f_2}{f_1^2+{f_2^2}/{2}}, 
\end{equation}   
being an upper bound for the generation of SSC at a fixed temperature, see green dashed-line of Fig.~\ref{normCoheQuQu}. In Eq.~\eqref{cohe_max}, $\mathcal{C}_0$ represents the $l_1$-norm of coherence of the target qubit
at the steady-state, at zero temperature and for $N$ being large, 
i.e., $\mathcal{C}_0$ 
is $T\rightarrow 0$ and
$N\rightarrow\infty$ limit of~(\ref{cohe_ss_qq}).
Notice that $\mathcal{C}_0$ as a function of $f_1$ or $f_2$ reaches 
an upper bound of
${1}/{\sqrt{2}}\approx 0.7$
for $f_2=\sqrt{2}f_1$. To our best knowledge, and based on numerical evidence, this upper bound represents an absolute maximum of coherence attainable within models assumed in our work.
This complements the recent numerical results obtained in~\cite{guarnieri2020non}, where the authors found that, only for $N=1$, the maximal amount of SSC  is achievable when the weights of the parallel and orthogonal components, of a repeated interaction similar to Eq.~(\ref{hami_int_qubit_qubit}), are equal to each other.
Let us recall that Eq.~(\ref{hami_int_qubit_qubit}) can be rewritten as $\mathcal{V}_I=f_1\sigma_z\otimes\sigma_x^B +f_2(\sigma_x\otimes\sigma_x^B+\sigma_y\otimes\sigma_y^B)/2$. In comparison, the interaction considered in~\cite{guarnieri2020non} is, in our notation, $\mathcal{V}_I=J_{zy}\sigma_z\otimes\sigma_y^B+(J_x\sigma_x\otimes\sigma_x^B+J_y\sigma_y\otimes\sigma_y^B)$, where each $J_i$ is a real coupling constant, see Eq.~(30) of~\cite{guarnieri2020non}. There, the authors found that the combination $J_{zy}=J_x=J_y$ gives the maximal amount of SSC. 

It is instructive to know how close $\mathcal{C}_{\rm ss}^\infty$ can be to the ideal situation where the qubit is in a pure coherent superposition of its two energy eigenstates. First note that the state vector $|\psi\rangle$ for such coherent superposition can be written as $|\psi\rangle=(|e\rangle+|g\rangle)/\sqrt{2}$, where $|e\rangle$ ($|g\rangle$) is the excited (ground) state. The corresponding density matrix $\rho=|\psi\rangle\langle\psi|$ allows to use \textcolor{black}{$\mathcal{C}(t)$} and get an $l_1$-norm of coherence $\mathcal{C}=1.0$, which is the largest value of $\mathcal{C}$ that one can obtain for  a two-level system. In comparison, the maximum value of $\mathcal{C}_{\rm ss}^\infty$, as discussed in the previous paragraph, is $1/\sqrt{2}\approx 0.7$, i.e., nearly $70\%$ of the ideal situation. This means that, for the collision model described in Fig.~\ref{setup_clusters}, it is enough to have incoherent clusters made of a few qubits to generate a considerable amount of SSC (see Fig.~\ref{normCoheQuQu}).

To study quantum coherence in the target qubit we
have chosen the energy basis of the system as our 
preferred basis. However, from the above results 
we have no indication of the qubit final state purity. The purity represents, in principle, the coherence with respect to an optimally chosen basis (achieved by a proper change of the basis) which is instructive to compare with \textcolor{black}{$\mathcal{C}(t)$}. Hence, to characterize the final qubit state better, we calculate the purity $\mathcal{P}(t)={\rm tr}\{\rho^2(t)\}$~\cite{nielsen2002quantum}
which is a basis-independent quantity. The purity takes its maximum value $\mathcal{P}=1$ if the state is pure 
and its minimum of $\mathcal{P}=1/d$, with $d$ the dimension of the corresponding Hilbert space, when the state is completely mixed~\cite{jaeger_2010}. 
For the simplest case of a density matrix of a qubit, 
the purity in the steady state can be easily written as
$\mathcal{P}_{\rm ss}=(1+\mathcal{C}_{\rm ss}^2+\langle\sigma_z\rangle_{\rm ss}^2)/2$, where
$\langle\sigma_z\rangle_{\rm ss}$
 and $\mathcal{C}_{\rm ss}$ are defined in~(\ref{sigma_z_ss}) of the Appendix and~(\ref{cohe_ss_qq}), respectively.

The inset of Fig.~\ref{normCoheQuQu} shows the behaviour of steady-state purity $\mathcal{P}_{\rm ss}$ as a function of the scaled temperature of the bath qubits. We point out that for low temperatures, the final state of 
the target qubit is close to a pure state, especially when it interacts only with one bath qubit at a time, see Fig.~\ref{normCoheQuQu} solid blue line ($N=1$). In contrast to the $l_1$-norm of coherence $\mathcal{C}_{\rm ss}$, the purity $\mathcal{P}_{\rm ss}$ decreases with the number of qubits 
$N$ in each bath cluster. In particular, for $N\gg 1$ the purity in the steady state is well approximated by
\begin{equation}\label{purity_approx}
\mathcal{P}_{\rm ss}\approx\frac{1}{2}+\Big(\frac{1}{2}+\frac{f_2^2}{8f_1^2}\Big)\mathcal{C}_0^2\tanh^2(\beta\omega_B/2).
\end{equation}
When the second term of the above equation vanishes at high temperatures, the purity $\mathcal{P}_{\rm ss}$ reduces to its minimum value of $1/2$, i.e., the final state of the target qubit is a completely mixed state, see blue dashed line in the inset of Fig.~\ref{normCoheQuQu}. For any other finite value of $N$ the purity $\mathcal{P}_{\rm ss}$ will fall between these two limiting curves, dashed-blue and top solid-blue, Fig.~\ref{normCoheQuQu}. Moreover, the plateau region in $\mathcal{C}_{\rm ss}$ and $\mathcal{P}_{\rm ss}$ allows for
reaching their maximum values in the limit  $0< k_BT/\hbar\omega_B\ll 1$. In such a low-temperature regime, the purity decreases from its maximum value as proportional to $\mathcal{C}_0^2$, i.e., is of the second order in the generated maximum coherence $\mathcal{C}_0$.
Remarkably, we notice a trade-off between coherence and purity for generation of SSC. In particular, relation
(\ref{purity_approx}) shows that the purity is a quadratic 
function of the SSC, cf. with Eq.~\eqref{cohe_max}.

\textcolor{black}{Physically, the cluster schemes discussed above allow for more significant system coherence, as they effectively increase the system-bath coupling strength~\cite{Ustinov_2021,Wang_PRL_2021}. This is due to the additive nature of the interaction, see Eq.~\eqref{ham:int:qubit:cluster}. Such stronger coupling, in turn, causes stronger system correlations with the clusters, resulting in a lower final system-state purity, see Fig.~\ref{normCoheQuQu}}.

It is interesting to see how the energy population of the target qubit in the steady state, measured by $\langle \sigma_z\rangle_{\rm ss}$, is modified due to the generation of SSC. For instance, when $N\gg 1$, such expectation value is well approximated by $\langle\sigma_z\rangle_{\rm ss}\approx-\tanh(\beta\omega_B/2)[1-{f_1}\mathcal{C}_0/f_2]$, see Eq.~(\ref{sigma_z_ss}) for its exact value. This result shows that the formation of SSC induces corrections in the thermal population of the target qubit with respect to the case when the system is coupled  to an effective thermal bath solely via RWA interaction, i.e., if $f_1=0$, $f_2\neq 0$. In such case, the qubit population inversion is equal to the standard Boltzmann factor $2n_F-1=-\tanh(\beta\omega_B/2)$, where $n_F=\big(\exp\big[\hbar\omega_B/(k_BT)\big]+1\big)^{-1}$ is the Fermi-Dirac mean occupation number of the spin system. Notice that these corrections to population are of the same order of magnitude as SSC, because they depend on $f_1^2$ and $f_2^2$. These type of corrections were recently point it out in~\cite{Purkayastha2020} and in the supplementary material of~\cite{SSC_Radim}. 
We additionally remark that, to get such corrections in those works, a quite complex perturbation expansion of a generalized equilibrium state has to be used for the derivation, in contrast to the  simple calculations presented in our work. 

So far, we have considered solely the RWA type of interaction, as in Eqs.~(\ref{hami_int_qubit_qubit}) 
and~(\ref{ham:int:qubit:cluster}), including only rotating (RWA) terms in $\mathcal{V}_I$. However, if we want to examine the possible effects of the counter-rotating (C-R) terms included in the qubit system and the bath elements interaction Hamiltonian, we should use, e.g., the form
\begin{equation}\label{hami_int_qubit_rabi_type}
\mathcal{V}_I=f_1\sigma_z\otimes(\sigma_-^B+\sigma_+^B)
+f_2(\sigma_-+\sigma_+)\otimes(\sigma_-^B+\sigma_+^B),
\end{equation}
which can be rewritten as 
$\mathcal{V}_I=f_1\sigma_z\otimes\sigma_x^B+f_2\sigma_x\otimes\sigma_x^B$.
This energy non-preserving interaction contains the~C-R terms $\sigma_+\otimes\sigma_+^B$ and 
$\sigma_-\otimes\sigma_-^B$, that were neglected 
in the second term of Eq.~(\ref{hami_int_qubit_qubit}), reflecting the use of RWA. Importantly, for interaction~(\ref{hami_int_qubit_rabi_type}),
we can identify the parallel and orthogonal projections $\mathcal{H}^{\parallel  \mathcal{H}_{S}}=f_1\sigma_z\otimes\sigma_x^B$ and $\mathcal{H}^{\perp \mathcal{H}_{S}}=f_2\sigma_x\otimes\sigma_x^B$
respectively, again in the spirit of \cite{SSC_Radim}. 
With respect to  the discussion in section~\ref{sec_qubit_ho}, we may alternatively refer to Eq.~(\ref{hami_int_qubit_rabi_type}) as the Rabi-type interaction Hamiltonian. Such interaction is available in both trapped ions and superconducting circuit experiments \cite{FluhmannNature2019,campagneibarcq2019quantum}. 
The interaction Hamiltonian~(\ref{hami_int_qubit_rabi_type}) can be also rewritten as $\mathcal{V}_I=s^\dagger A+s A^\dagger$ with $s=f_1\sigma_z+f_2\sigma_x$ and $A=\sigma_-^B$. Here, $s$ is a Hermitian operator, $s=s^\dagger$.  Therefore, using the 
interaction~(\ref{hami_int_qubit_rabi_type}) in (\ref{familiar_Lindblad_2}), the following master equation can be derived
\begin{equation}\label{eq:mast:rabi:q}
\frac{d\rho}{dt}=-\frac{i\omega}{2}[\sigma_z,\rho]+\langle\{\sigma_-^B,\sigma_+^B\}\rangle\mathcal{L}[f_1\sigma_z+f_2\sigma_x]\rho.
\end{equation}
In Appendix~\ref{apx_bloch_rabi} we show that it is not possible to generate SSC in the target qubit if the dynamics is described by the above master equation \eqref{eq:mast:rabi:q}. By comparison with Eq.~\eqref{eq:mast:2} we recognize their similar structure up to the  term $f_2\sigma_+$ neglected in the definition of   the argument $s=f_1\sigma_z+f_2\sigma_-$ of the Lindbladian Eq.~\eqref{eq:mast:2} with respect to Eq.~\eqref{eq:mast:rabi:q}. We can intuitively understand \eqref{eq:mast:rabi:q} as infinite bath temperature limit of Eq.~\eqref{eq:mast:2}, due to the equality $\langle\{\sigma_-^B,\sigma_+^B\}\rangle/2=\lim_{T\rightarrow\infty}\langle\sigma_-^B\sigma_+^B\rangle =\lim_{T\rightarrow\infty}\langle\sigma_+^B\sigma_-^B\rangle = 1/2$. Thus, the system dynamics determined by Eq.~\eqref{eq:mast:rabi:q} will generate no  steady-state coherence in the energy basis of the system, as it might be interpreted similarly as $\lim_{T\rightarrow\infty}\mathcal{C}_{\rm ss}$, see Eq.~\eqref{cohe_ss}, and
this is vanishing as $T^{-1}$ in the high temperature limit (see discussion below Eq.~\eqref{s_function}). 

However, it is quite remarkable that coherence in the energy basis of the target qubit can still be generated during the time evolution, see next subsection~\ref{TLS-transient} and also the end of Appendix~\ref{apx_bloch_rabi} for details.
As a matter of fact, for $f_1\cdot f_2=0$ Eq.~(\ref{eq:mast:rabi:q}) will not generate coherence, even in the transient evolution.

\subsection{Optimal Transient State Coherence}
\label{TLS-transient}
In the situations considered in section~\ref{TLS-steady}, we have examined the qubit system properties in the ``long time'' (many collisions) limit. For the tests using pulsed experimental control \cite{FluhmannNature2019,campagneibarcq2019quantum} it is advantageous to know, if the quantum coherence is attainable in the transient state  (TS) (finite time/number of collisions) regime. Such question was possible to ask in the previous work \cite{SSC_Radim} as well, but there it was an extremely complex task to answer it, compared to the collision interactions used here.  Therefore, we present here the results for the maximum value (with respect to time/number of collisions) of coherence, $\overline{\mathcal{C}}^{{\rm TLS}}_{{\rm TS}}$, attainable after some specific, {\it optimized}, time/number of collisions, for otherwise fixed values of the rest of the parameters, and its comparison to the asymptotic value of SSC. The results presented below are good approximation of the exact numerical solution in the regime of small (with respect to system frequency) values of system-bath (TLS clusters in this case) coupling constants $f_{1(2)}$ (specified below).

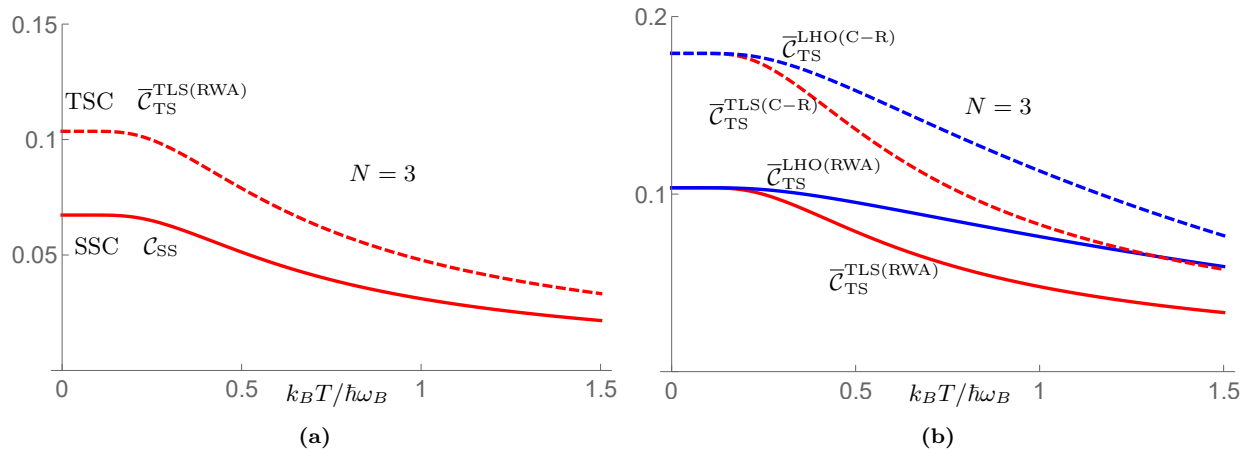
\begin{figure*}[t]
\subfloat[\label{fig-TSC-SSC-comparison}]{\begin{tikzpicture} 
  \node (img1)  {\includegraphics[width=.93\columnwidth]{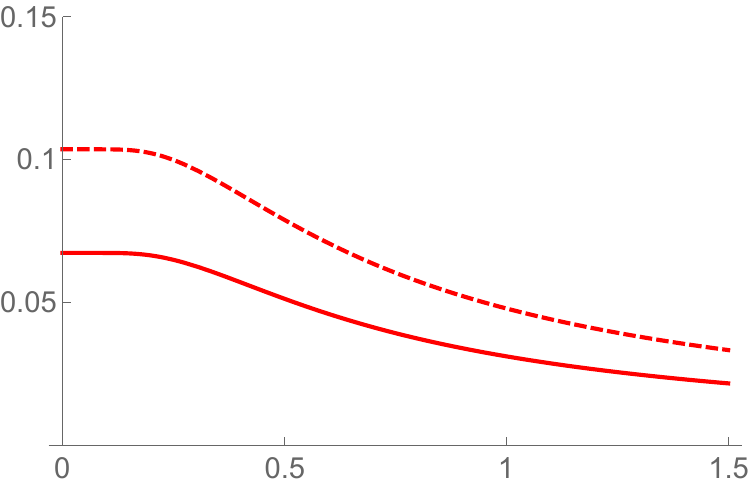}};
  \node[above=of img1, node distance=0cm, yshift=-6.5cm,xshift=0.3cm] {$k_BT/\hbar\omega_B$};
  \node[above=of img1, node distance=0cm, yshift=-4.5cm,xshift=-2.5cm] {SSC$\quad {\mathcal{C}}_{\rm SS}$};
  \node[above=of img1, node distance=0cm, yshift=-2.6cm,xshift=-2.1cm] {TSC$\quad \overline{\mathcal{C}}^{{\rm TLS(RWA)}}_{{\rm TS}}$};
  \node[above=of img1, node distance=0cm, yshift=-3.5cm,xshift=.9cm] {$N=3$};
\end{tikzpicture}}
\subfloat[\label{fig-TLS-LHO-comparison}]{\begin{tikzpicture} 
  \node (img1)  {\includegraphics[width=.93\columnwidth]{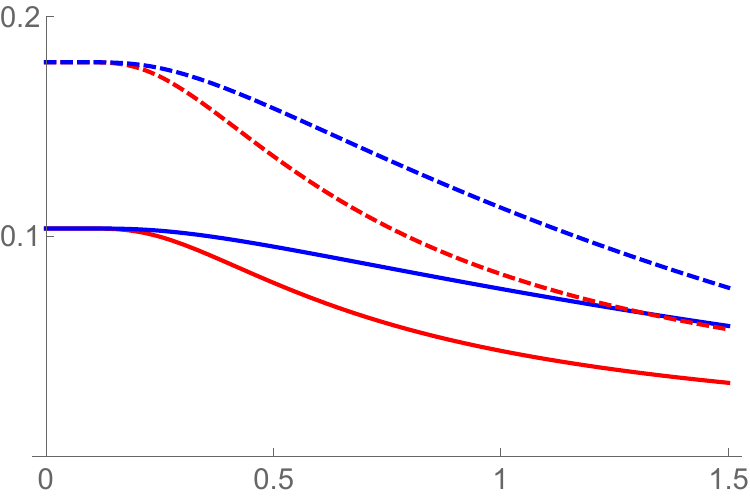}};
  \node[above=of img1, node distance=0cm, yshift=-6.6cm,xshift=0.3cm] {$k_BT/\hbar\omega_B$};
  \node[above=of img1, node distance=0cm, yshift=-5.1cm,xshift=-.7cm] {$\overline{\mathcal{C}}^{{\rm TLS(RWA)}}_{{\rm TS}}$};
  \node[above=of img1, node distance=0cm, yshift=-3.7cm,xshift=-1.5cm] {$\overline{\mathcal{C}}^{{\rm LHO(RWA)}}_{{\rm TS}}$};
  \node[above=of img1, node distance=0cm, yshift=-2.9cm,xshift=-2.3cm] {$\overline{\mathcal{C}}^{{\rm TLS(C-R)}}_{{\rm TS}}$};
  \node[above=of img1, node distance=0cm, yshift=-2.cm,xshift=-1.3cm] {$\overline{\mathcal{C}}^{{\rm LHO(C-R)}}_{{\rm TS}}$};
  \node[above=of img1, node distance=0cm, yshift=-2.7cm,xshift=.8cm] {$N=3$};
\end{tikzpicture}}
\caption{{\bf (a)} Comparison of the achievable coherence of the qubit interacting with clusters of $N=3$ bath TLS and its dependence on temperature $T$. The figure shows the superiority and the typical behavior of the optimized TSC, Eq.~\eqref{eq-trans-coh-approx-TLS}, over the SSC, Eq.~\eqref{cohe_ss_qq}, for the same values of the relevant parameters, $f_1=f_2=0.15$, $\omega=\omega_B=1$, $N=3$ in the small-to-moderate $f_{1(2)}$ regime. {\bf (b)} The comparison of the optimized coherence achievable in the transient regime (TSC) with the linear harmonic oscillator (LHO) and two level system (TLS) clusters with $N=3$ bath units and its dependence on the bath temperature $T$. The oscillator (LHO) clearly shows an advantage over the two level systems (TLS) in yielding higher TSC for the same values of the parameters $f_1=f_2=0.15$, $\omega=\omega_B=1$, $N=3$, c.f. Eqs.~\eqref{eq-trans-coh-approx-LHO} and \eqref{eq-trans-coh-approx-TLS}.
}
\end{figure*}

As the coherence in the transient regime is determined by the system dynamics, one has to workout the solution of the corresponding Bloch Eqs.~(\ref{bloch_JCM}) and~(\ref{bloch_Rabi}). These apply to the RWA approximated (RWA) interaction, cf. Eq.~(\ref{ham:int:qubit:cluster}), and interaction including counter-rotating (C-R) terms, cf. Eq.~(\ref{hami_int_qubit_rabi_type}), respectively. 

These solutions read formally
\begin{equation}
\begin{split}
\langle\vec{\sigma}\rangle^{{\rm RWA}}&=\exp[{\bf B}\,t](\langle \vec{\sigma}\rangle_0+{\bf B}^{-1}  \vec{c})-{\bf B}^{-1} \vec{c},\\
\langle\vec{\sigma}\rangle^{{\rm C-R}}&=\exp[\mathbb{B}\,t]\langle \vec{\sigma}\rangle_0,
\end{split}
\end{equation}
where $\langle \vec{\sigma}\rangle=(\langle\sigma_x\rangle,\langle\sigma_y\rangle,\langle\sigma_z\rangle)^T$ and $\langle \vec{\sigma}\rangle_0=(0,0,z_0)^T$ stands for the initial Bloch vector with $z_0=-\tanh(\beta\omega/2)$ the inversion of the system thermal population, chosen as a natural initial condition. The superscripts RWA (C-R) reflect the type of interaction between the system qubit and the bath TLS cluster, resulting in different differential-system defining matrices ${\bf B}$ ($\mathbb{B}$), and constant vector $\vec{c}$, defined in Eqs.~(\ref{bloch_JCM}) and~(\ref{bloch_Rabi}), respectively.

The derivation of the time-optimized values of coherence and purity is based on approximate solution of the above mentioned Bloch equations. The solution can be found, e.g., by the Laplace transform method and assuming small enough damping terms in the corresponding Bloch equations, see Appendix~\ref{apx_bloch} for more details. The resulting optimal values of transient state coherence (TSC) for a qubit colliding with clusters of qubits (of the size $N$) read 
\begin{eqnarray}\nonumber
\overline{\mathcal{C}}^{{\rm TLS(RWA)}}_{{\rm TS}}&\approx & 2f_1f_2N|z_0|\exp[-\pi N(f_1^2+f_2^2/4)/\omega]/\omega,\\
\overline{\mathcal{C}}^{{\rm TLS(C-R)}}_{{\rm TS}}&\approx & 4f_1f_2N|z_0|\exp[-\pi N(f_1^2+f_2^2)/\omega]/\omega,
\label{eq-trans-coh-approx-TLS}
\end{eqnarray}
being a good approximation of the exact numerical results, if the parameters satisfy $f_1,\, f_2\ll \omega=\omega_B$, $N\lessapprox 3$, and $1/2 \lessapprox |z_0|\leq 1$,
see horizontal gray dashed line of Fig.~\ref{bloch_evolution}, as a typical example. As one can note, the transient state (TS) coherence scales with $f_1$, $f_2$ in the same way as its steady-state counterpart $\mathcal{C}_{\rm ss}$ in the low-temperature and weak-interaction limits. 
The derivation of Eqs.~\eqref{eq-trans-coh-approx-TLS} (and \eqref{eq-trans-purity-TLS} below) assumes that the initial system state is in thermal equilibrium with the bath and that the system is resonant with each bath element. In the limit of weak damping assumed here, the time at which the system coherence is maximized reads $t_{\rm max}\approx \pi/\omega$, cf. Fig.~\ref{bloch_evolution} for an example.

In the same regime of parameters, we obtain results for the optimised system purity $\overline{\mathcal{P}}$ (in the same time instant as Eq.~\eqref{eq-trans-coh-approx-TLS}). They read
\begin{equation}
\begin{split}
\overline{\mathcal{P}}^{{\rm TLS(RWA)}}_{{\rm TS}}&\approx \frac{1}{2}\left(1+z_0^2\right),\label{eq-trans-purity-TLS}\\
\overline{\mathcal{P}}^{{\rm TLS(C-R)}}_{{\rm TS}}&\approx \frac{1}{2}\left(1+z_0^2\exp[-2\pi N(f_1^2+f_2^2)/\omega]\right).
\end{split}
\end{equation}
The regime of parameter values, in which the above approximations work well, describes effectively the underdamped dynamics, in the sense that the effective system damping being weak enough. In the opposite overdamped case, one should better resort to numerical evaluation.

Comparing the above results, we can see the surprising effect of the counter rotating (C-R) terms $(\sigma_+\otimes\sigma_+^B+h.c.)$ on the optimized system purity $\overline{\mathcal{P}}^{{\rm TLS(C-R)}}_{{\rm TS}}$ with respect to the RWA approximated $\overline{\mathcal{P}}^{{\rm TLS(RWA)}}_{{\rm TS}}$. These terms boost the thermally generated transient state coherence $\overline{\mathcal{C}}^{{\rm TLS(C-R)}}_{{\rm TS}}$. The counter rotating terms result in additional basis sensitive quantum correlation of the system and the bath elements, lowering the system purity noticeably, even for the relatively short evolution times, but at the same time creating larger off-diagonal terms in the system state, see Fig.~\ref{fig-TLS-cohmax}. Within the validity range of Eqs.~\eqref{eq-trans-coh-approx-TLS} and \eqref{eq-trans-purity-TLS}, limited by the values $f_1,\, f_2\ll \omega=\omega_B$, $N\lessapprox 3$, and $1/2 \lessapprox |z_0|\leq 1$, we can see the trend, cf. Fig.~\ref{fig-TLS-cohmax} for $N=\{1,3\}$, showing that increasing the number $N$ of the cluster units leads to increase of the coherence $\overline{\mathcal{C}}_{\rm TS}$ and decrease of the corresponding purity $\overline{\mathcal{P}}_{\rm TS}$, provided the rest of the parameters being fixed ($f_{1(2)}\approx 0.15$). Such feature generally holds for both types of interactions, i.e., RWA or C-R. Thorough numerical investigation of more precise quantitative behavior of the quantities of interest is beyond the scope of this paper. 

In general, focusing on the TSC can be more profitable compared to SSC. The first positive aspect is the smaller number of interactions (shorter waiting time) necessary to reach the respective coherence value. This should be understood as follows. Microscopically, every real (pulsed) or numerical experiment consists of certain number of interactions (collisions) between the system of interest and the (bath) units. Each such interaction, has some (possibly small, but) finite duration $\tau$. Thus, the total evolution time is proportional to the number of collisions $t=n\tau$. Such numerical experiment underlies our work as well, see Fig.~\ref{bloch_evolution}, although we prefer to use the effective master equation reasoning, predominantly. 

Another positive aspect of TSC is that it is larger than SSC, being certainly true for C-R case, where SSC even vanishes, see discussion below Eq.~\eqref{eq:mast:rabi:q}. In the case of RWA interaction, the situation is a little bit more complex. In the small to moderate $f_{1(2)}$ values regime, c.f. Eq.~\eqref{eq-trans-coh-approx-TLS}, the TSC value always overcomes the SSC, c.f. Fig.~\ref{fig-TSC-SSC-comparison}. In the regime of strong system-bath coupling, e.g., for the parameters' values used in Fig.~\ref{normCoheQuQu}, the time-optimized value of the coherence coincides with the SSC, Eq.~\eqref{cohe_ss_qq}.
\section{SSC from oscillator bath elements}\label{sec_qubit_ho}
\subsection{Asymptotic Coherence}
\label{LHO-steady}
\begin{figure}[t]
\includegraphics[width=.95\columnwidth]{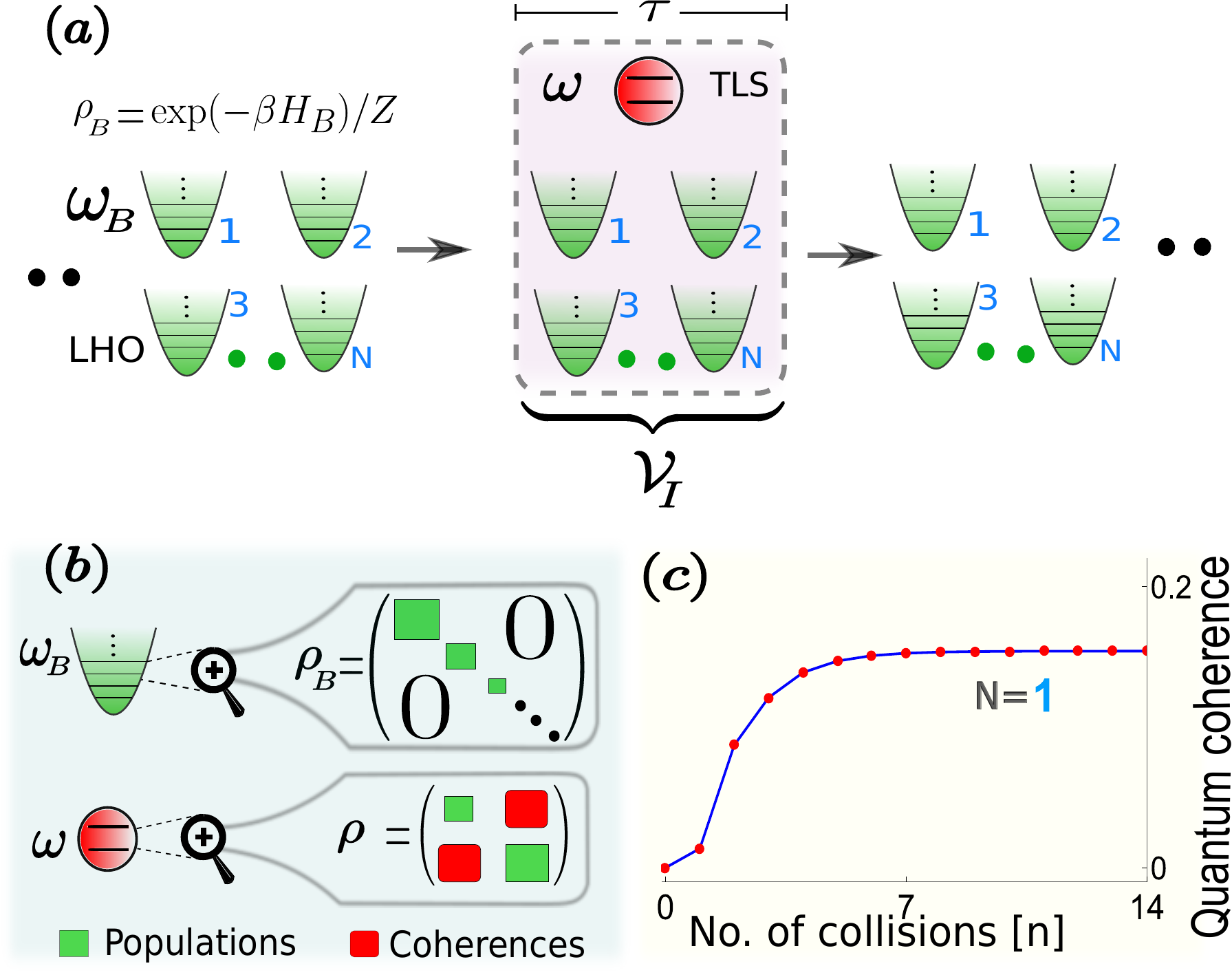} 
\caption{
{\bf (a)} Schematic showing clusters of $N$ independent and non-correlated linear harmonic oscillators (LHO) of frequency $\omega_B$ as the bath elements (green harmonic potentials) replacing the bath qubits, cf. Fig.~\ref{setup_clusters}. {\bf (b)} Before their interaction with the target qubit  (red) of frequency $\omega$, the LHO are initially in a thermal state $\rho_B$ where its populations (green squares) follow the standard Boltzmann distribution. After successive interactions, the density matrix $\rho$ of the target qubit has coherence (red squares). {\bf (c)} Typical behavior of the generation of SSC in the target qubit for the simplest case of $N=1$ bath LHO in each cluster.}
\label{setup}
\end{figure}

The oscillators forming the bath can carry more coherence than qubits, therefore, it might be fruitful to consider an oscillator bath to generate SSC or TSC. Thus, by choosing particular composite system-bath interaction~\cite{SSC_Radim}
\begin{eqnarray}\label{hami_int_qubit}
\mathcal{V}_I=f_1\sigma_z\otimes(b+b^\dagger)+f_2(\sigma_+\otimes b+\sigma_-\otimes b^\dagger),
\end{eqnarray}
we assume the bath elements to be 
linear harmonic oscillators (LHO), instead of the two-level baths of previous section, see Fig.~\ref{setup}. Similarly as in the previous section~\ref{TLS-steady}, the coherence is at first generated in the bath, then transferred back to the system. But in the LHO case, the bath cluster can not be saturated (due to the infinite dimension) and thus coherence could be expected larger, in principle, if the interaction strength $f_1$ increases.  
Formally, the interaction~(\ref{hami_int_qubit}) is obtained from (\ref{hami_int_qubit_qubit}), by replacing $\sigma_-^B$ ($\sigma_+^B$) with $b$ ($b^\dagger$), where $b$ ($b^\dagger$) 
is the annihilation (creation) operator of the quantum LHO, obeying the standard commutation relation $[b,b^\dagger]=1$. We observe that 
(\ref{hami_int_qubit}) can be rewritten as 
$\mathcal{V}_I=s\otimes A^\dagger+s^\dagger\otimes A$ if $s=f_1\sigma_z+f_2\sigma_-$ and $A=b$.
Therefore, it is straightforward to show that the reduced 
dynamics of the target qubit will be described by equation identical to (\ref{eq:mast:2}), with the only difference that 
we need to replace $\langle\sigma_-^B\sigma_+^B\rangle$ 
by $\langle bb^\dagger\rangle$ and $\langle\sigma_+^B\sigma_-^B\rangle$
by $\langle b^\dagger b\rangle$.
As in the previous section, we assume here, that each LHO is in a thermal state such that 
$\langle b^\dagger b\rangle=n_T$,
$\langle\{b,b^\dagger\}\rangle=2n_T+1$ and 
$\langle[b,b^\dagger]\rangle=1$. Here, $n_T$ is the average Bose-Einstein occupation number given by
$n_T=\big(\exp\big[{\hbar\omega_B}/({k_BT})\big]-1\big)^{-1}$,
whereas at high temperatures $n_T\sim k_BT/(\hbar\omega_B)$.

The interaction~(\ref{hami_int_qubit}) is easily generalized,
as in the previous section, to the case in which the target qubit interacts {\em collectively} with bath clusters made of $N$ non-correlated and independent harmonic oscillators. For such case, the corresponding master equation describing the target qubit dynamics and its $l_1$-norm of coherence in the steady-state are, basically, the same as the results~(\ref{eq:mast:2}) and~(\ref{cohe_ss}), respectively, with the only difference that the expressions in~(\ref{conm_rel_spins}) must be substituted by their bosonic counterparts
\begin{align}\label{conm_rel_bosons}
\langle\{B,B^\dagger\}\rangle=N\coth(\beta\omega_B/2),\quad
\langle[B,B^\dagger]\rangle=N.
\end{align}
We have defined $B\equiv\sum_{k=1}^N b_k$ and $B^\dagger\equiv\sum_{k=1}^N b_k^\dagger$ as the collective
annihilation and creation bath operators of each cluster,
respectively. 

\begin{figure*}
\subfloat[\label{normCohe2}]{\begin{tikzpicture} 
  \node (img1)  {\includegraphics[width=.93\columnwidth]{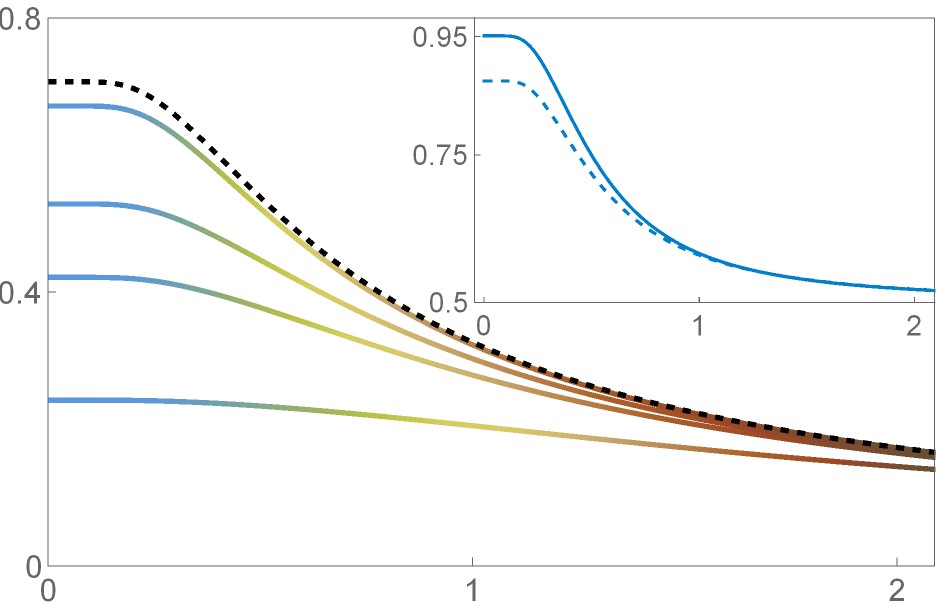}};
  \node[above=of img1, node distance=0cm, yshift=-6.8cm,xshift=0cm] {$k_BT/\hbar\omega_B$};
  \node[above=of img1, node distance=0cm, yshift=-3.3cm,xshift=-3.1cm] {$N=3$};
  \node[above=of img1, node distance=0cm, yshift=-2.6cm,xshift=-3.1cm] {$N=8$};
  \node[above=of img1, node distance=0cm, yshift=-3.5cm,xshift=.8cm] {$N\gg 1$};
  \node[above=of img1, node distance=0cm, yshift=-1.8cm,xshift=-3.1cm] {$N\gg 1$};
  \node[above=of img1, node distance=0cm, yshift=-4.1cm,xshift=-3.1cm] {$N=2$};
  \node[above=of img1, node distance=0cm, yshift=-5.2cm,xshift=-3.1cm] {$N=1$};
   \node[above=of img1, node distance=0cm, yshift=-2.cm,xshift=1.3cm] {$N=1$};
   \node[above=of img1, node distance=0cm, yshift=-2.1cm,xshift=-1.3cm] {RWA interaction};
  \node[above=of img1, node distance=0cm, yshift=-1.2cm,xshift=.2cm] {$\mathcal{P}_{\rm ss}$};
   \node[above=of img1, node distance=0cm, yshift=-1.2cm,xshift=-3.1cm] {{\color{black}${\mathcal{C}}_{\rm ss}$}};
\end{tikzpicture}}
\subfloat[\label{fig-LHO-cohmax}]{\begin{tikzpicture} 
  \node (img1)  {\includegraphics[width=.93\columnwidth]{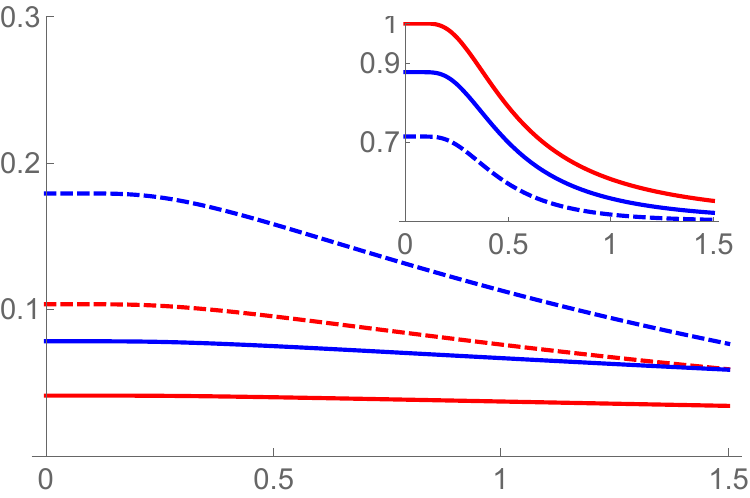}};
  \node[above=of img1, node distance=0cm, yshift=-6.8cm,xshift=0cm] {$k_BT/\hbar\omega_B$};
  \node[above=of img1, node distance=0cm, yshift=-3.2cm,xshift=-2.7cm] {$N_{\rm C-R}=3$};
  \node[above=of img1, node distance=0cm,rotate=-30,anchor=center, yshift=-2.1cm,xshift=2.5cm] {$N=3$};
  \node[above=of img1, node distance=0cm, yshift=-4.3cm,xshift=-2.7cm] {$N_{\rm RWA}=3$};
  \node[above=of img1, node distance=0cm, yshift=-5.3cm,xshift=-2.7cm] {$N_{\rm C-R}=1$};
  \node[above=of img1, node distance=0cm,rotate=-30,anchor=center, yshift=-1.6cm,xshift=1.9cm] {$N=1$};
  \node[above=of img1, node distance=0cm, yshift=-5.8cm,xshift=-2.7cm] {$N_{\rm RWA}=1$};
   \node[above=of img1, node distance=0cm, yshift=-2.cm,xshift=1.9cm] {$N=\{1,3\}$};
  \node[above=of img1, node distance=0cm, yshift=-1.4cm,xshift=.8cm] {$\overline{\mathcal{P}}_{\rm TS}^{\rm LHO}$};
  \node[above=of img1, node distance=0cm, yshift=-3.5cm,xshift=.8cm] {{\color{blue}C-R}};
  \node[above=of img1, node distance=0cm, yshift=-2.7cm,xshift=2.25cm] {{\color{red}RWA}};
   \node[above=of img1, node distance=0cm, yshift=-1.5cm,xshift=-3.cm] {{\color{black}${\overline{\mathcal{C}}}_{\rm TS}^{\rm LHO}$}};
\end{tikzpicture}}
\caption{{\bf (a)} Steady-state coherence (SSC) with respect to the energy basis of the qubit system repeatedly colliding with the stream of LHO (representing the bath), cf. Fig.~\ref{setup}. As above, the coherence is quantified with the $l_1$--norm of coherence, \textcolor{black}{$\mathcal{C}(t)$}, as a function of the bath oscillator's scaled temperature, see lowest solid line ($N=1$). SSC can increase substantially if the qubit interacts with clusters of $N$ non-correlated and independent harmonic oscillators, see upper black lines. The parameters are $\omega=\omega_B=1$, 
$f_1=f_2/\sqrt{2}$, $f_2=0.6$ and $N=\{1,2,3,8\}$ 
(solid lines). The dashed-line corresponds to the 
theoretical limit~(\ref{cohe_max}) for 
$N\gg 1$. {\bf (b)} The dependence of time-optimized coherence, Eq.~\eqref{eq-trans-coh-approx-LHO}, on the bath temperature in cases when the system interacts with $N=\{1,3\}$ bath LHO via RWA interaction (labeled RWA), Eq.~\eqref{hami_int_qubit}, or with counter-rotating (C-R) terms included, Eq.~\eqref{hami_int_qubit_rabi}. As in the case of TLS bath, the (C-R) results clearly have an edge over the (RWA) results in terms of attainable coherence $\overline{\mathcal{C}}$. Similarly, the corresponding system state purity $\overline{\mathcal{P}}$, Eqs.~\eqref{eq-trans-purity-LHO}, of (C-R) interaction is suppressed with respect to the (RWA) scenario. The temperature dependence is entering the results through the system initial inversion $z_0=-\tanh(\hbar\omega/(2k_BT))$ and assumption that the system and the bath have initially the same temperature $T$ and are resonant $\omega=\omega_B$. The values of the parameters are the same as in Fig.~\ref{fig-TLS-cohmax}. Please note the different scales on the vertical axes on panels {\bf (a)} and {\bf (b)}, reasons being the same as in Fig.~\ref{fig-TLS-cohmax}.}
\end{figure*}

To obtain the $l_1$-norm of coherence in the target qubit, the expectation values~(\ref{conm_rel_bosons}) have to be 
used in Eqs.~(\ref{r_function}) and~(\ref{s_function}). 
We point out that an increase of steady-state coherence, as a function of the number of bath LHO in each cluster, is possible, see solid lines of Fig.~\ref{normCohe2}. It is important to mention that, although the overall behavior of the quantities plotted in Fig.~\ref{normCoheQuQu} and Fig.~\ref{normCohe2} has similar form, it differs in details. For instance, in Fig.~\ref{normCohe2} the decrease of SSC with the bath temperature is slower compared to the behavior plotted in Fig.~\ref{normCoheQuQu}. However, when the number of oscillators $N$ within the clusters is large, the $l_1$-norm of coherence in the steady-state reduces
to $\mathcal{C}_{\rm ss}=\mathcal{C}_0\tanh(\beta\omega_B/2)$,
which is the same limit found in previous section~\ref{TLS-steady},
see black dashed-line of Fig.~\ref{normCohe2}.
This result can be understood in the following way, with a clear link to the results of section~\ref{TLS-steady}. When the number $N$ of TLSs bath elements increases, 
one can always use the Holstein-Primakoff representation~\cite{Ban93} 
in which the collective spin operators $S_\pm$, \eqref{ham:int:qubit:cluster}, can be written as bosonic operators in such a way that the interaction~(\ref{ham:int:qubit:cluster}) and the  expression of collective interaction~(\ref{hami_int_qubit})
become equivalent. This procedure is sometimes called the thermodynamic limit~\cite{HirschJ}, meaning that $N\rightarrow\infty$. Mathematically, this is know as the Heisenberg-Weyl contraction of the $SU(2)$ Lie group.
On the other hand, if each bath element (either harmonic oscillator or qubit) is prepared in its ground state (i.e., at zero temperature) then both $\coth(\beta\omega_B/2)$ and $\tanh(\beta\omega_B/2)$ approach unity, hence, the expectation values (\ref{conm_rel_bosons}) and (\ref{conm_rel_spins}) are the same. This means that at low temperatures the target qubit reaches the same SSC values regardless if the stream of 
bath elements is made of harmonic oscillators or a 
set of qubits. This could have been anticipated, because at low enough temperatures, each harmonic oscillator behaves as an effective two-level system due to the fact that there are not enough thermal excitation to populate more than the first excited state. We can therefore advantageously use bath oscillators to extended experimental platforms suitable for the tests and to obtain SSC for larger temperatures. However, for large enough bath temperatures, the $\mathcal{C}_{\rm ss}$ scales approximately as $\omega_B/T$, approaching zero as in the case of TLS bath clusters. 

The inset of Fig.~\ref{normCohe2} shows the 
steady-state purity $\mathcal{P}_{\rm ss}=(\langle\sigma_z\rangle_{\rm ss}^2+\mathcal{C}_{\rm ss}^2+1)/2$ of the target qubit as a function of the scaled temperature $k_BT/\hbar\omega_B$ for two limit cases, when the bath clusters are  made of one harmonic oscillator (blue solid-line) and when these contain a large number $N\gg 1$ of harmonic oscillators (blue dashed-line). 
From the above-mentioned argument we know that the explicit 
expression of $\mathcal{P}_{\rm ss}$, for $N\gg 1$, is given by~(\ref{purity_approx}). Contrary to the $l_1$-norm of coherence, the purity decays faster in this configuration, compared to the case with bath qubit. This observation can be made from careful comparison of the corresponding insets of Fig.~\ref{normCoheQuQu} 
and Fig.~\ref{normCohe2}.
It confirms the already described trade-off between the SSC and purity which rise a benchmark for further investigation of SSC.

Alternatively, we may take into account the counter-rotating terms in~(\ref{hami_int_qubit}), obtaining 
\begin{align}\label{hami_int_qubit_rabi}
\mathcal{V}_I=&f_1\sigma_z\otimes(b+b^\dagger)+f_2(\sigma_-+\sigma_+)\otimes(b+b^\dagger),
\end{align}
which resembles the interaction 
Hamiltonian~(\ref{hami_int_qubit_rabi_type}).
The second term of~(\ref{hami_int_qubit_rabi}) is known as the
quantum Rabi interaction, which is often written as~$ \sigma_x\otimes X_b$~\cite{PolRMP19},
with $X_b\equiv b+b^\dagger$. The quantum Rabi interaction
describes, in the fields of cavity and circuit quantum
electrodynamics, the ultra-strong coupling regime 
between the electromagnetic radiation and matter at its 
most fundamental level~\cite{FriskKockum2019}. To study principal appearance of SSC using the trapped ion experiments, it can be induced in a controllable way by two-tone external drive \cite{FluhmannNature2019}.
It is easy to show that the corresponding master equation 
of the reduced dynamics for a target qubit, describing an
interaction like~(\ref{hami_int_qubit_rabi}), will be given
by~(\ref{eq:mast:rabi:q}) with the replacement 
$\langle\{\sigma_-^B,\sigma_+^B\}\rangle\rightarrow\langle\{b,b^\dagger\}\rangle$. Therefore, no SSC can be created. The coherence occurs only during the transient dynamics governed by this master equation. 

\subsection{Optimised Transient State Characteristics}
\label{LHO-transient}

As in the previous section, we compare the value of coherence and state purity generated in the steady state with the time-optimised values possible to acquire during the transient. These results reflect the experimental possibility to interrupt the target system evolution at certain point. We assume small values $f_{1(2)}$, system and bath elements to be resonant, and both in thermal initial state at temperature $T$. These assumptions yield the approximate values of coherence maxima (with respect to the time) as
\begin{eqnarray}\nonumber
\overline{\mathcal{C}}^{{\rm LHO(RWA)}}_{{\rm TS}}&\approx & 2f_1f_2N\exp[-\pi N(f_1^2+f_2^2/4)/(|z_0|\omega)]/\omega,\\
\overline{\mathcal{C}}^{{\rm LHO(C-R)}}_{{\rm TS}}&\approx & 4f_1f_2N\exp[-\pi N(f_1^2+f_2^2)/(|z_0|\omega)]/\omega,
\label{eq-trans-coh-approx-LHO}
\end{eqnarray}
well corresponding to the numerical results if the parameters satisfy $f_1,\, f_2\ll \omega=\omega_B$, $N\lessapprox 3$, and $1/2 \lessapprox |z_0|\leq 1$, c.f. Fig.~\ref{bloch_evolution}. As in the case of TLS from the previous section, we note that the transient state coherence (TSC) has the same scaling as its steady state (SSC) counterpart in the low-temperature and weak (in the same sense as in previous sections) coupling limits. 

In the same range of parameters, we can derive the values of purity achievable at the same instant of evolution as in Eq.~\eqref{eq-trans-coh-approx-LHO}, reading
\begin{eqnarray}
\overline{\mathcal{P}}^{{\rm LHO(RWA)}}_{{\rm TS}}&\approx &\frac{1}{2}\left(1+z_0^2\right),\label{eq-trans-purity-LHO}\\
\nonumber
\overline{\mathcal{P}}^{{\rm LHO(C-R)}}_{{\rm TS}}&\approx &\frac{1}{2}\left(1+z_0^2\exp[-2\pi N(f_1^2+f_2^2)/(|z_0|\omega)]\right).
\end{eqnarray}
This difference is basically resulting from the effect of the counter rotating terms ($\sigma_+\otimes b^\dagger$+h.c.) in the interaction Hamiltonian, present for the Rabi interaction, see Fig.~\ref{fig-LHO-cohmax}. 

As in the previous section, see \ref{TLS-transient} for details, the general comparison of the coherence achievable in TSC vs. SSC regime remains the same. The typical behavior of the system coherence when using the LHO clusters is qualitatively the same as in Fig.~\ref{fig-TSC-SSC-comparison}, TSC being superior to SSC values of coherence for the same parameters in the RWA moderate interaction regime, the same being true for the C-R interaction. 

At the end of this subsection, we would like to compare the results for coherence achievable with clusters of LHO vs. TLS bath units, again in the moderate coupling regime. Comparison of the results stemming from Eq.~\eqref{eq-trans-coh-approx-TLS} and \eqref{eq-trans-coh-approx-LHO} shows the superiority of LHO over the TLS bath units in generating the TSC coherence, see Fig.~\ref{fig-TLS-LHO-comparison}. These results point at the role of the bath units dimension in generation of the coherence in the transient dynamics and that the higher dimension of the units might be preferable for reaching higher TSC values. More thorough analysis and comparison in the strong coupling regime should rely on fully numerical approach beyond the scope of this paper. It will be useful for the preparation of proof-of-principle experiment with trapped ions or superconducting circuits.

\section{SSC in a target oscillator}\label{SSC_oscillator}

So far, we have focused on generating SSC in two-level
target systems. Due to the essential role of quantum 
systems with infinite Hilbert space, as quantum mechanical resonators, in the development of 
quantum technologies with hybrid systems~\cite{Kurizki3866}, 
in this section, we would like to point out the analysis 
of replacing the target qubit from previous sections
with a quantum harmonic oscillator with free Hamiltonian 
$\mathcal{H}_S=\omega_0 a^\dagger a$. 
Remarkably, we have found cases where the 
composite system-bath repeated interaction, having 
counter-rotating terms can generate SSC in the target harmonic oscillator as well. 
From the resource theory approach of quantum thermodynamics, this is also interesting because it has been recently proven~\cite{PRL_Muller_2019} that only reference frames (systems displaying quantum coherence) with infinite Hilbert space can be used to perform ``catalytic coherence''~\cite{PRL_Aberg_2014}, a weaker form of coherence broadcasting, see also~\cite{PRL_Iman_2019}.  
For instance, 
$\mathcal{V}_I=f_1 a^\dagger a\otimes X_b+f_2X_a\otimes X_b$
represents the case of the target oscillator
interacting with a bath oscillator, where $X_c\equiv c+c^\dagger$. 
Notice that $\mathcal{V}_I$ can also be written as the bi-linear 
combination $s^\dagger\otimes A+s\otimes A^\dagger$ with 
$s=f_1 a^\dagger a+f_2 X_a$ and $A=b$. Following the 
procedure described in previous sections and in the 
Appendix~\ref{model}, it is easy to show that 
the corresponding master equation for the target
oscillator is given by
\begin{equation}\label{mas_eq_osc}
\frac{d\rho}{dt}=-{i\omega_0}[a^\dagger a,\rho]+\langle\{b,b^\dagger\}\rangle\mathcal{L}[f_1 a^\dagger a+f_2 X_a]\rho,
\end{equation}
which resembles Eq.~(\ref{eq:mast:rabi:q}). 
The corresponding expectation value $\langle a\rangle$ at the steady state 
is (see Appendix~\ref{eqs_oscillaor_case}):
$\langle a \rangle_{\rm ss}=-f_1f_2/(f_1^2+i2\tilde{\omega})$,
where $\tilde{\omega}\equiv \omega_0/\langle \{b,b^\dagger\}\rangle$,
causing a displacement of the target oscillator by the quantity
$\langle X_a\rangle_{\rm ss}=-2f_1f_2/(f_1^2+4\tilde{\omega}^2f_1^{-2})$.
This result shows that the target oscillator will end up in a steady 
state with some degree of coherence, independently of its initial state,
as long as the product $f_1\cdot f_2$ is nonzero. Recall, that for any incoherent state $\rho_{\rm inc}$ of the harmonic 
oscillator, ${\rm tr}\{\rho_{\rm inc}a\}\equiv \langle a\rangle_{\rm inc}=0$. 
Remarkably, there is a stark difference between qualitative properties of the results, if using LHO instead of TLS as a target system, as long as the bath units are LHOs interacting mutually by C-R type of interaction, see sec.~\ref{LHO-steady}. The LHO system acquires nonzero SSC in the $T\rightarrow 0$ limit, in contrast to TLS system, and this SSC {\it survives} even in the high temperature limit.

When the bath oscillator is replaced by clusters of $N$ non-correlated 
and independent harmonic oscillators, we just need to change 
$\langle\{b,b^\dagger\}\rangle$ by $\langle\{B,B^\dagger\}\rangle$ in
Eq.~(\ref{mas_eq_osc}) and in the expression of $\langle a\rangle_{\rm ss}$. 

Another interesting composite interaction in the spirit of Eq.~\eqref{hami_int_qubit} is
$\mathcal{V}_I=f_1 a^\dagger a\otimes(b+b^\dagger)+f_2(a^\dagger\otimes b+a\otimes b^\dagger)$,
where we recognize the first term of $\mathcal{V}_I$ as an 
opto-mechanical interaction and the second one as the usual 
coupling between two harmonic oscillators with the RWA.
Within the repeated interactions approach the corresponding 
master equation of the target oscillator is
\begin{equation}\label{eq_maes_osc_rwa}
\begin{split}
\frac{d\rho}{dt}=-i\omega_0[a^\dagger a,\rho]+\langle bb^\dagger\rangle\mathcal{L}[f_1a^\dagger a+f_2a]\rho\qquad \\
+\langle b^\dagger b\rangle\mathcal{L}[f_1a^\dagger a+f_2a^\dagger]\rho,
\end{split}
\end{equation}
notice the similarity with Eq.~(\ref{eq:mast:2}). 
In Appendix~\ref{eqs_oscillaor_case}, we have found some approximate results suggesting that Eq.~(\ref{eq_maes_osc_rwa}) could generate SSC in the target harmonic oscillator. Additionally, we require the bath oscillators to be in a thermal state with a temperature $T\neq 0$, as the thermal population $n_T$ stands in the nominator of $\langle a \rangle_{\rm ss}$. Therefore, no SSC can be generated in the target oscillator at low temperatures. This example contrasts with what was found in previous sections, where the SSC is maximum precisely at $T=0$. Similarly to the previous RWA case, Eq.~\eqref{mas_eq_osc}, the nonzero average $\langle a \rangle_{\rm ss}$ appears in the high $T$ limit of the bath oscillators.

Similar results can be obtained for two-level systems in thermal 
states replace the bath oscillators in the two previous interactions. 
For such case, the non-energy preserving interaction is, for example, 
$\mathcal{V}_I=f_1 a^\dagger a\otimes\sigma_x^B+f_2X_a\otimes\sigma_x^B$,
which represents the inverse scenario of Fig.~\ref{setup}, where 
the role of the target system and the bath elements is interchanged.

The results of this short section underline the strong dimension-dependent differences, influencing SSC attainability in the target system, depending on the bath units nature.

\section{Thermodynamic cost for the generation of SSC}\label{therm_cost}

In any collision model there is an implicit time dependence present in the microscopic switching-on and switching-off of the interaction between the target system and the bath elements. 
From this perspective, any collision model is microscopically non autonomous, although it can be effectively described by a master equation (time evolution) corresponding to an autonomous system. This can be useful for current experimental platforms to simulate the appearance of SSC. 
The implicit time dependence means that from a quantum thermodynamic point of view, one can consider the corresponding thermodynamic cost for such process~\cite{Barra2015,Esposito_PRX_17}. 

In this section, we investigate this cost using the appropriate expressions for the heat and work necessary to maintain the steady state and thus SSC. To do that, we will use the general formulas for heat $\dot{Q}$, work $\dot{W}$ and internal energy ${\rm d}\langle \mathcal{H}_S\rangle /{\rm d}t$ rates derived in the Appendix B of~\cite{De_Chiara_2018} for a boundary-driven Lindblad master equations, like the ones used throughout this work. These formulas, in our notation, yield
\begin{subequations}\label{aprop_express}
\begin{eqnarray}
\dot{Q}&=&\frac{1}{2}\langle [\mathcal{V}_I,[\mathcal{V}_I,\mathcal{H}_B]]\rangle,
\\
\dot{W}&=&-\frac{1}{2}\langle
[\mathcal{V}_I,[\mathcal{V}_I,\mathcal{H}_S+\mathcal{H}_B]]\rangle,
\\
\frac{d\langle \mathcal{H}_S\rangle}{dt}&=&-\frac{1}{2}\langle [\mathcal{V}_I,[\mathcal{V}_I,\mathcal{H}_S]]\rangle,
\end{eqnarray}
\end{subequations}
where $\mathcal{H}_S$ ($\mathcal{H}_B$) is the system (bath element) free Hamiltonian. The expectation values in the above equations have to be calculated with respect to product state  $\rho(t)\otimes\rho_B$, where $\rho_B$ is the initial incoherent (thermal) state of each bath element and $\rho(t)$ is the instantaneous state of the target system. The quantities in Eqs.~\eqref{aprop_express} satisfy the dynamical version of the First Law~\cite{KoslDyna} ${d\langle \mathcal{H}_S\rangle}/{dt}=\dot{Q}+\dot{W}$. 
With these definitions, the $\dot{Q}$, $\dot{W}$ quantities correspond to the following sign convention: if they are {\it injected} (added) to the system, they are {\it positive}.  

We first consider the situation studied in Sec.~\ref{TLS-steady}, where the bath elements are two-level systems, i.e., $\mathcal{H}_B=\omega_B\sigma_z^B/2$ and $\mathcal{V}_I$ is given in Eq.~(\ref{hami_int_qubit_qubit}). For such case it is easy to show that
\begin{subequations}\label{therm_express_TLS}
\begin{align}
\dot{Q}&=f_2^2\omega_B\big(n_F-\langle\sigma_+\sigma_-\rangle\big)+f_1^2\omega_B(2n_F-1)\nonumber\\
&\quad+f_1f_2\omega_B\langle\sigma_x\rangle,\label{heat_TLS_baths}\\
\dot{W}&=(\omega-\omega_B)f_2^2\big(n_F-\langle\sigma_+\sigma_-\rangle\big)-f_1^2\omega_B(2n_F-1)\nonumber\\
&\quad+(\omega-2\omega_B)f_1f_2\langle\sigma_x\rangle/2,\label{work_TLS_baths}\\
\frac{d\langle \mathcal{H}_S\rangle}{dt} &=f_2^2\omega\big(n_F-\langle\sigma_+\sigma_-\rangle\big)+\frac{1}{2}f_1f_2\omega\langle\sigma_x\rangle,\label{U_TLS_baths}
\end{align}
\end{subequations}
where $n_F=(\exp[\hbar\omega_B/(k_BT)]+1)^{-1}$. In consistency with the rest of the paper, we consider the resonant case $\omega =\omega_B$ in the following three paragraphs. Furthermore, for the sake of simplicity, we focus on the case $N=1$ (single bath unit at each interaction) and to the steady state situation to avoid, e.g., the initial state ambiguities. 

In such steady-state settings, the LHS of \eqref{U_TLS_baths} is zero, yielding the expected energy balance $\dot{W}_{\rm ss}=-\dot{Q}_{\rm ss}$. In the case of energy conserving interaction, $f_1=0$, Eqs.~\eqref{work_TLS_baths}, \eqref{heat_TLS_baths} dictate ${d\langle \mathcal{H}_S\rangle}_{\rm ss}/{dt}=\dot{W}_{\rm ss}=\dot{Q}_{\rm ss}\equiv 0$, i.e., true equilibrium situation with the energy currents vanishing. On contrary, $f_1\cdot f_2\neq 0$ regime leads the system to a {\it non-equilibrium} steady state, characterized by a non-vanishing power input and dissipated heat rate $\dot{W}_{\rm ss}=-\dot{Q}_{\rm ss}>0$, describing the energy cost per collision, ${W}_{\rm ss}=\dot{W}_{\rm ss}\tau>0$, maintaining the steady state, ($\tau$ being the collision duration). This injected energy is directly dissipated to the bath as heat ${Q}_{\rm ss}=-\dot{W}_{\rm ss}\tau<0$, per collision. 

Explicitly, we obtain for $\mathcal{C}_{\rm ss}> 0$
\begin{equation}
    \dot{W}_{\rm ss}(T)=\frac{|f_1|\;\omega\;(2\omega^2+4f_1^4+f_1^2f_2^2)}{|f_2|\sqrt{4\omega^2+(4f_1^2+f_2^2)^2}}\;\mathcal{C}_{\rm ss}(T)> 0,
    \label{eq-wdot-steady-tls}
\end{equation}
where $\mathcal{C}_{\rm ss}$ is the SSC, see Eq.~\eqref{cohe_ss}, showing proportionality between the power input and achieved energy basis coherence of the system TLS, balanced by the heat flowing to the bath during each collision. Such type of relation, Eq.~\eqref{eq-wdot-steady-tls}, opens the possibilities for optimization (minimization) of $\dot{W}_{\rm ss}$ for a given achieved SSC.

Here we also want to emphasize the non-trivial splitting in terms of heat and work of Eq.~(\ref{U_TLS_baths}). Unlike in Ref.~\cite{De_Chiara_2018}, knowing the change of the system's internal energy only, obtained from the Lindblad master equation, is {\it insufficient} to identify the appropriate expressions of heat (\ref{heat_TLS_baths}) and work (\ref{work_TLS_baths}) involved in our collision model using composite interactions, showing richer thermodynamic aspects of obtaining the energetic coherence. This fact can be underlined by another observation regarding the interactions described by Eq.~\eqref{hami_int_qubit_rabi_type}. While with this interaction we can not obtain SSC, the work input rate is strictly positive, $\dot{W}_{\rm ss}=\omega(f_1^2+f_2^2)\tanh{(\beta\omega/2)}\geq 0$, even in this case, moreover being larger than \eqref{eq-wdot-steady-tls} for the same relevant parameters. This sets up a possibility of broader thermodynamic evaluation, as from these examples it is clear that discussing the thermodynamic efficiency of obtaining SSC can be nontrivial.

The second example corresponds to the case of harmonic oscillators bath elements (see Sec.~\ref{LHO-steady}) with $\mathcal{H}_B=\omega_B b^\dagger b$, $\mathcal{V}_I$ being given by Eq.~(\ref{hami_int_qubit}). In this case the appropriate thermodynamic expressions (\ref{aprop_express}) are:
\begin{subequations}\label{therm_express_HO}
\begin{align}
\dot{Q}&=f_2^2\omega_B\big[n_T-(2n_T+1)\langle\sigma_+\sigma_-\rangle\big]-f_1^2\omega_B\nonumber\\
&\qquad\quad+f_1f_2\omega_B(2n_T+1)\langle\sigma_x\rangle,\label{heat_Ho_baths}\\
\dot{W}&=(\omega-\omega_B)f_2^2\big[n_T-(2n_T+1)\langle\sigma_+\sigma_-\rangle\big]+f_1^2\omega_B\nonumber\\
&\qquad\quad+(\omega-2\omega_B)f_1f_2(2n_T+1)\langle\sigma_x\rangle/2,\label{work_HO_baths}\\
\frac{d\langle \mathcal{H}_S\rangle}{dt} &=f_2^2\omega[n_T-(2n_T+1)\langle\sigma_+\sigma_-\rangle]\nonumber\\
&\qquad\qquad\qquad+f_1f_2\omega(2n_T+1)\langle\sigma_x\rangle/2,\label{U_HO_baths}
\end{align}
\end{subequations}
where $n_T=(\exp[{\hbar\omega_B}/({k_BT})]-1)^{-1}$. Remarkably, one can easily show that Eqs.~(\ref{therm_express_TLS}) and Eqs.~(\ref{therm_express_HO}) coincide when the bath elements are prepared in the ground state. This happens because $n_{F,T}$ vanishes for $T\rightarrow 0$ and means that in addition to reaching the same SSC in the target system, either with qubits or harmonic oscillators as the bath elements, the associated thermodynamic cost is also the same in the low temperature regime. On the other hand, at high temperatures and resonant conditions, the steady-solution of, for example Eq.~(\ref{work_HO_baths}), is simply $\dot{W}_{\rm ss}\approx \omega f_1^4/(f_1^2+f_2^2/2)$.  In general, for $f_1\cdot f_2\neq 0$ $\dot{W}$ is nonzero because $[\mathcal{V}_I,\mathcal{H}_S+\mathcal{H}_B]\neq 0$, no matter if the resonance condition $\omega=\omega_B$ is satisfied. Therefore, the generation of SSC using composite interactions is an out-of-equilibrium situation, accompanied by a work cost given in Eqs.~(\ref{work_TLS_baths}) and (\ref{work_HO_baths}) for qubits, respectively, and oscillators bath elements. Similar conclusions were obtained in~\cite{guarnieri2020non} but using a purely numerical calculation. In contrast, our analytical results help us to better understand each quantity's thermodynamic  structure in relatively simple terms. For example, one can verify that in the steady state and resonant conditions, $\dot{W}_{\rm ss}$ in Eq.~(\ref{work_HO_baths}) is  proportional to the SSC through $\langle \sigma_x\rangle_{\rm ss}$, see Eq.~(\ref{norm_coherence_ss}), as it is the case of TLS bath units, given in Eq.~\eqref{eq-wdot-steady-tls}.
\textcolor{black}{When $f_1=0$ in Eq.~(\ref{heat_TLS_baths}), this reduces to a Landauer-like expression $\dot{Q}=f_2^2\omega_B(n_F-\langle\sigma_+\sigma_-\rangle)$~\cite{Segal_NanoL_2020}. Therefore, the generation of SSC induces strong modifications in the heat current that deviate from the well-known formulation of transport theory.}

Finally, we would like to discuss some implications of the above results on our collision model's physical interpretation. The fact that the work cost is nonzero (due to the on-off switching of the interaction) during the SSC generation, makes the model, strictly speaking, non-autonomous. This is a consequence of the particular time-dependent system-bath interaction we have chosen. However, it is worth noting that the corresponding Lindblad description in the continuous-time limit regime, which we also consider, clearly describes the dynamics of a non-driven and time-independent, i.e., autonomous, open quantum system, see e.g. Eq.~(\ref{eq:mast:2}). In this sense, we can gradually approach description of autonomous dynamics by a sequence of short time non-autonomous collisions. It can be useful to simulate such effects for many current experimental platforms with trapped ions or superconducting circuits.

\section{Conclusions}\label{sec:conclusions}

We have shown that quantum coherence can be 
generated, together with high purity, on a target quantum system when this shortly and repeatedly interacts with individual bath elements initially in incoherent (thermal) states. 
In our collision approach, typical for many experimental platforms with trapped cold ions and superconducting circuits, a large number of bath elements plays the role of an entire bath. These models not only give theoretical insight to the  microscopic processes in the baths required to achieve SSC, but mainly, they can be straightforwardly implemented using current experimental platforms.

Similarly to the previous work \cite{SSC_Radim}, we confirm here that the composite nature of the system-bath interaction represents an important condition to obtain the system  steady-state coherence (SSC). However, our results clearly show that modelling the open-systems by spin-boson and collision models is {\it not fully equivalent}. For the spin-boson model the composite interaction represents a sufficient condition \cite{SSC_Radim}, i.e., {\it if} the interaction is composite, {\it then} SSC is created. On contrary, our collision based results reveal that in such case of modelling, the composite interaction represents a necessary condition on SSC only, i.e., {\it if} one wants to create SSC, {\it then} the interaction has to be composite.

We stress that, unlike coherence trapping~\cite{SabrinaPRA2014}, the SSC is 
independent of the initial state of the target system.  Moreover, SSC can be created on individual quantum system. This is because, from the point of view of the bath, the compound target system does not need to have parts indistinguishable like in~\cite{LatunePRA2019,LatunePRR2019}. Such SSC can be increased substantially, if collective interactions between the target system and clusters of bath units are introduced. We observe that for low temperatures of the bath, the amount of SSC does not depend on the exact nature of the bath elements, both qubit or oscillator baths units reaching the same SSC value. For higher temperatures, the SSC is however higher for the oscillators-composed baths. This is of practical importance because one might have flexibility in choosing which physical systems best fit the experimental needs.


Due to the simple dynamics generated in our collision model allowing insight to the microscopic processes during the interaction, we have been able to study the generation of the transient coherence (TSC) in the regime of weak-to-intermediate values of the system bath coupling constants $f_{1(2)}$ in an approximate manner. Within this interaction regime we have found that for a wide range of parameters 
optimized TSC surpasses SSC, especially in the low-temperature regime. Moreover, in the TSC regime, it is more profitable to employ oscillator bath units than two-level system units, as well, as the former generate higher coherence of the target system. 

Remarkably, the simple structure of our results also allows to characterize
the intimate relationship between steady-state or transient coherence and
the state purity. In particular, we have found that, for a given interaction
Hamiltonian of the composite form [with parallel and orthogonal component of the interaction, see discussion below Eq.~(\ref{hami_int_qubit_qubit})], the coherence and purity reach their maximum in presence of a zero
temperature bath, possessing a small constant plateau in the regime of
small temperature (thus allowing for experimental observation) and finally show
monotonic decrease for increasing temperature.
It is furthermore worth noticing, however, that coherence and purity
behave in the opposite way with respect to the presence of counter-rotating
terms in the interaction Hamiltonian: for every fixed temperature $T$, the
lack of these terms leads to an {\it increase} in the maximum achievable purity and a
{\it decrease} in the corresponding maximum amount of coherence (where by this
we mean the maximum of the SSC within the RWA approximation compared to the
maximum of the TSC when counter-rotating terms are present).

Although our results show positive effects in the sense of generating relatively high SSC or TSC and purity, one may naturally ask if these results represent any fundamental limits. The answer is negative, thus the way how to beat the maximum coherence values achieved within models and settings assumed in our work can be a good future research target. Our results are of course based on our assumed models and the properties of used states, e.g., states of the bath units. Thus, if we would relax some requirements/assumptions on the bath-state properties, we might speculate on the  increase of SSC and TSC values {\it and jointly} the system purity. Another way leading to possibly overcoming the limits of our current models may lie in search for 
more effective (in terms of coherence generation) Hamiltonians and protocols, or, e.g., in the extension of the system-bath interaction time. Such modification brings the evolution beyond the one described in our work, namely to a more complex one including terms of higher order than linear in the interaction time. Full analysis of such possible scenarios is definitely suitable topic for future work.

From the thermodynamic perspective, we have considered the power input necessary to maintain the non-equilibrium steady state featuring SSC. Although this has been done under relatively simplifying assumptions (resonance, single bath unit interaction only), we have clearly established direct connection between the inevitable energy input and the produced SSC, showing direct proportionality between these quantities. On contrary, we have recognized system-bath interactions, which also need nonzero power input to evolve the system to the steady state, while {\it not} generating any SSC, hence the power input being unused, in a sense. Thus, we have characterized the steady state from different, and in fact complementary, perspectives, which might be stimulating for further and deeper analysis of SSC from a thermodynamic point of view.

It may be noted that, while following from \cite{SSC_Radim}, in our present analysis we have focused on interactions of the system with a single bath described by the classes of system-bath interaction Hamiltonians of the form $H_{\rm int}
= \sum_j O_{s,j} \otimes b_E + h.c.$ The more general type $H_{\rm int} = \sum_j
O_{s,j} \otimes b_{E,j} + h.c.$, where the summation index is extended
to the bath operators, can be considered as well (although, still describing the interaction with a single bath). Composite interactions belonging
to the latter and not included in the former can, in certain cases, also lead to the
generation of SSC, a recent example of which was considered in a
qubit-based collision model in Eq. (30) of \cite{guarnieri2020non}, where the presence of
counter-rotating terms also allowed for the observation of SSC. While being 
beyond the scope of the present work, this represents an interesting
outlook for future work.

\begin{acknowledgments}

R.R.-A. wants to thanks Professor \"Ozg\"ur E. 
M\"ustecapl{\i}o\u{g}lu for his hospitality 
at  Ko\c{c} University where the initial part of this work was done. R.F. and M.K. gratefully acknowledge support through Projects No. 20-16577S of the Czech Science Foundation and LTAUSA19099 from the Czech Ministry of Education, Youth and Sports. G.G. acknowledges support from FQXi and DFG FOR 2724 and also from the European Union Horizon 2020 research and innovation programme under the Marie Skłodowska-Curie Grant Agreement No.
101026667.
\end{acknowledgments}

\appendix

\section{The collision model of the system-bath interaction}
\label{model}
This Appendix describes a simple and general collision model (see an example in Fig.~\ref{setup_clusters}). 
This consists of  the system of our main 
interest repeatedly interacting with a stream
of bath elements that are initially prepared in an incoherent state, namely the thermal state. As we show at the end of this section, a large number of the bath elements will play the role of an environment.
%
During the short time of interaction of duration $\tau$, the total Hamiltonian is 
\begin{equation}\label{total_hami}
\mathcal{H}=\mathcal{H}_S+\mathcal{H}_B+{\mathcal{V}_I}/{\sqrt{\tau}},
\end{equation}
where $\mathcal{H}_S$ and $\mathcal{H}_B$ are, respectively, the system free Hamiltonian and the free Hamiltonian of one of the bath elements, and $\mathcal{V}_I$ represents the interaction between these two. 
Note that, for mathematical reasons that will become clear bellow, we have rescaled the interaction term by a factor $1/\sqrt{\tau}$ \cite{KarevskiPRL09,LandiPRE2014,Esposito_PRX_17,De_Chiara_2018,rodrigues2019,guarnieri2020non}. 
Apparently, shorter $\tau$ increases the interaction energy. 

Further, we assume that each bath element {\em before} its 
interaction with the system of interest at a time $t_n=n\tau$,
does not share any correlation with the latter 
and with any other bath element, so the state of the total system, $\rho_{\rm tot}(t_n)$, is given by the tensor product between the system state denoted by $\rho(t_n)$, and a thermal state $\rho_B$ of the incoming bath element:
$\rho_{\rm tot}(t_n)=\rho(n\tau)\otimes\rho_B$, 
where $\rho_B=\exp(-\beta\mathcal{H}_B)Z^{-1}$, 
$Z$ is the partition function and $\beta\equiv(k_BT)^{-1}$ the inverse scaled-temperature.
{\em After} the interaction with a bath element, the state of the system of interest at time $t_{n+1}$ is given by the stroboscopic map~\cite{rodrigues2019}:
%
$\rho[(n+1)\tau]={\rm tr}_B\{\rho_{\rm tot}'(t_{n+1})\}\equiv 
{\rm tr }_B\{U[\rho(n\tau)\otimes\rho_B]{U}^\dagger\}$,
%
where $U=\exp(-i\mathcal{H}\tau)$ is the
evolution operator of the total system and
${\rm tr}_B$ is the partial trace over the
bath degrees of freedom. We can use
the Baker–Campbell–Hausdorff formula to 
compute the unitary transformation up to the second order in $\tau$
\begin{equation}\label{unitary_transf}
\begin{split}
\rho_{\rm tot}'(t_{n+1})&= e^{-i\mathcal{H}\tau}\rho(n\tau)\otimes\rho_B e^{i\mathcal{H}\tau}\\
&=\rho(n\tau)\otimes\rho_B-[i\mathcal{H}\tau,\rho(n\tau)\otimes\rho_B]\\
&+\frac{1}{2!}[i\mathcal{H}\tau,[i\mathcal{H}\tau,\rho(n\tau)\otimes\rho_B]]+O(\tau^3),
\end{split}
\end{equation}
which after using~(\ref{total_hami}) 
in~(\ref{unitary_transf}) and keeping terms 
at most linear in $\tau$ yields
\begin{equation}
\begin{split}
\rho_{\rm tot}'(t_{n+1}&)=
-i\tau[\mathcal{H}_S+\mathcal{H}_B+{\mathcal{V}_I}/{\sqrt{\tau}},\rho(n\tau)\otimes\rho_B]\\
&-\frac{\tau}{2}[\mathcal{V}_I,[\mathcal{V}_I,\rho(n\tau)\otimes\rho_B]]+\rho(n\tau)\otimes\rho_B.
\end{split}
\end{equation}
Taking the partial trace over the bath $B$ in the above expression and without any loss of generality assuming ${\rm tr}_B\{\mathcal{V}_I\rho_B\}=0$,
as customary~\cite{rodrigues2019,Palma_2012,Ciccarello17,Rivas2012}, we get 
\begin{equation}
\begin{split}\label{custumary}
\rho((n+1)\tau)-\rho(n&\tau)=-i\tau[\mathcal{H}_S,\rho(n\tau)]\\
&-\frac{\tau}{2}{\rm tr}_B\{[\mathcal{V}_I,[\mathcal{V}_I,\rho(n\tau)\otimes\rho_B]]\},
\end{split}
\end{equation}
which does not depend on the free bath Hamiltonian $\mathcal{H}_B$. 
The condition ${\rm tr_B}\{\mathcal{V}_I\rho_B\}=0$ does not restrict the interaction with the bath elements, actually, such assumption can be enforced by moving into the interaction picture representation of a rescaled local Hamiltonian of the system, see~\cite{Palma_2012,Rivas2012}. 
For a particular example where ${\rm tr_B}\{\mathcal{V}_I\rho_B\}\neq 0$ and its impact on the spectral response of the target system see~Ref.~\cite{roman2019spectral}. Then, the continuous-time limit of the model can be obtained if we  divide~(\ref{custumary}) by $\tau$ and
take the limit $\tau\rightarrow 0$~\cite{ciccarello2021quantum, ciccarello2017collision,Ciccarello13,Angsar19,Deniz19,roman2019spectral}. This yields
the reduced dynamics of the qubit density matrix 
as~\cite{De_Chiara_2018,Esposito_PRX_17,rodrigues2019} 
\begin{equation}\label{eq:mast:1}
\frac{{d}\rho}{{d}t}=-i[\mathcal{H}_S,\rho]-\frac{1}{2}{\rm tr}_B\Big\{\big[\mathcal{V}_I,[\mathcal{V}_I,\rho\otimes\rho_B]\big]\Big\},
\end{equation}
where $d\rho/dt\equiv\lim_{\tau\rightarrow 0}[\rho((n+1)\tau)-\rho(n\tau)]\tau^{-1}$.

For the case in which
$\mathcal{V}_I$ can be written as the bi-linear combination
$\mathcal{V}_I=s^\dagger A+sA^\dagger$ between system and bath operators, $s$ and $A$ respectively, the bath trace in~(\ref{eq:mast:1}) can be easily worked out. Thus, with such an interaction Hamiltonian, Eq.~(\ref{eq:mast:1}) acquires simple and more familiar Lindblad form:
\begin{equation}\label{familiar_Lindblad}
\frac{{d}\rho}{{d}t}=-i[\mathcal{H}_S,\rho]+\langle AA^\dagger\rangle\mathcal{L}[s]\rho+\langle A^\dagger A\rangle\mathcal{L}[s^\dagger]\rho, 
\end{equation}
where 
$\mathcal{L}[x]\rho\equiv x\rho x^\dagger-\frac{1}{2}(x^\dagger x\rho+\rho x^\dagger x)$ 
and $\langle x \rangle\equiv{\rm tr}\{x\rho_B\}$ with $\rho_B$ being the initial (thermal) state of the bath.
Using~(\ref{hami_int_qubit_qubit}) as the interaction Hamiltonian in~(\ref{familiar_Lindblad}) we obtain Eq.~(\ref{eq:mast:2}) of the main text. Let us point out that, for the special case in which $s$ is a Hermitian operator, $s=s^\dagger$,  Eq.~(\ref{familiar_Lindblad}) 
reduces to
\begin{equation}\label{familiar_Lindblad_2}
\frac{{d}\rho}{{d}t}=-i[\mathcal{H}_S,\rho]+\langle\{A,A^\dagger\}\rangle\mathcal{L}[s]\rho,
\end{equation}
where $\{x,x^\dagger\}=xx^\dagger+x^\dagger x$ is the anti-commutator. In section~\ref{TLS} we can see that~(\ref{familiar_Lindblad}) and~(\ref{familiar_Lindblad_2})
are useful master equations describing, respectively, system-bath interactions with and without
the rotating wave approximation.

\section{Bloch equations and steady sate coherence}\label{apx_bloch}

\subsection{Rotating-wave-approximated interactions}
\label{apx_bloch_jcm}

Here we describe how to derive equation~(\ref{cohe_ss}) of
the main text using the  interaction~(\ref{hami_int_qubit_qubit}) with the counter-rotating terms neglected (RWA performed). 
First, we should note that from Eq.~(\ref{eq:mast:2}) it is
easy to prove, after some algebra, the following identities:
\begin{subequations}
\begin{align}
{\rm tr }\big\{\mathcal{L}[f_1\sigma_z+f_2\sigma_\pm]\rho\,\sigma_x\big\}&=-\Big(2f_1^2+\frac{f_2^2}{2}\Big)\langle\sigma_x\rangle\nonumber\\ &\quad\,\,\,\,+f_1f_2\langle\sigma_z\rangle \mp 2 f_1f_2,\\
{\rm tr }\big\{\mathcal{L}[f_1\sigma_z+f_2\sigma_\pm]\rho\,\sigma_y\big\}&=-\Big(2f_1^2+\frac{f_2^2}{2}\Big)\langle\sigma_y\rangle,\\
{\rm tr }\big\{\mathcal{L}[f_1\sigma_z+f_2\sigma_\pm]\rho\,\sigma_z\big\}&=-f_2^2\langle\sigma_z\rangle+f_1f_2 \langle\sigma_x\rangle\pm f_2^2.
\end{align}
\end{subequations}
These identities will be useful to calculate the 
expectation values $\langle \sigma_i\rangle$, where 
$i=\{x,y,z\}$, with respect to the state $\rho$ 
of the target qubit. Defining the vectors  
$\vec{\sigma}=(\sigma_x,\sigma_y,\sigma_z)^{\intercal}$ 
and $\vec{c}=(c_x,0,-c_z)^{\intercal}$ and using
Eq.~(\ref{eq:mast:2}) along with the above expressions, 
the corresponding Bloch equations can be written as:
\begin{align}\label{bloch_JCM}
\frac{d}{dt}\langle\vec{\sigma}\rangle=\mathbf{B}\langle\vec{\sigma}\rangle+\vec{c},
\end{align}
where $\langle\vec{\sigma}\rangle$ 
is the Bloch vector and $\bf{B}$ is the following matrix
\begin{align}
\mathbf{B}=
\begin{pmatrix}
-\Gamma&-\omega&\Omega\\
\omega&-\Gamma&0\\
\Omega&0&-\gamma
\end{pmatrix}.
\end{align}
These Bloch equations follow directly from the quantum master equation \eqref{familiar_Lindblad}, without any further approximation or additional assumptions. 
We have defined the matrix elements of $\bf{B}$ as:
\begin{align}
\gamma &= f_2^2 \langle\{\sigma_-^B,\sigma_+^B\}\rangle,\qquad\quad\quad c_x =2(f_1/f_2)c_z,\\
\Omega &= f_1f_2 \langle\{\sigma_-^B,\sigma_+^B\}\rangle,\qquad\,\quad\, c_z = f_2^2\langle[\sigma_-^B,\sigma_+^B]\rangle,\\
\Gamma &=\big(2f_1^2+{f_2^2}/{2}\big)\langle\{\sigma_-^B,\sigma_+^B\}\rangle,\label{Gamma_eq}
\end{align}
with averaging done with respect to $\rho_B$, the initial (thermal) state of the bath.
Making $d\langle \vec{\sigma}\rangle/dt=0$ 
the steady-state values $\langle\sigma_i\rangle_{\rm ss}$ 
of (\ref{bloch_JCM}) are easily obtained:
\begin{align}
\langle \sigma_x\rangle_{\rm ss} &=\frac{f_1f_2\langle[\sigma_-^B,\sigma_+^B]\rangle\Gamma}{\Gamma^2+\omega^2-(f_1/f_2)^2\gamma\Gamma},\label{sig_x_ss}\\
\langle \sigma_y\rangle_{\rm ss} &=\frac{\omega}{\Gamma}\langle \sigma_x\rangle_{\rm ss}\label{sig_y_ss}, \\ 
\langle \sigma_z\rangle_{\rm ss} &=\frac{\Omega}{\gamma}\langle \sigma_x\rangle_{\rm ss}-\frac{c_z}{\gamma}.\label{sigma_z_ss}
\end{align}
To quantify the generation of SSC in the state 
$\rho$ of the target qubit we use the $l_1$-norm
of coherence, which is a suitable measure to 
compute it~\cite{PlenioCohe}.
For a two-level system this can be defined as 
$\mathcal{C}(t)=|\langle \sigma_x(t)\rangle+i\langle \sigma_y(t)\rangle|$.
At the steady state, and using (\ref{sig_y_ss}),
it reduces to
\begin{equation}\label{norm_coherence_ss}
\mathcal{C}_{\rm ss}=|\langle \sigma_x\rangle_{\rm ss}|\sqrt{1+\left(\frac{\omega}{\Gamma}\right)^2}.
\end{equation}
When we substitute (\ref{Gamma_eq}) and (\ref{sig_x_ss}) 
in the above expression we obtain Eq.~(\ref{cohe_ss}) of the 
main text. Evidently, all these results are easily generalized
for the case in which the the stream of bath single
qubits are replaced for a stream of bath clusters
that interact with the target qubit 
(see Fig.~\ref{setup_clusters}). In such case, we should
replace the commutator and ati-commutator for their 
respective expressions given by Eq.~(\ref{conm_rel_spins})
of the main text.

\subsection{Beyond-RWA interactions}\label{apx_bloch_rabi}

Here we derive the steady-state solution of the Bloch vector
when the master equation~(\ref{eq:mast:rabi:q}) of the main text is used to describe
the dynamics of the target qubit, i.e., when counter-rotating (C-R) terms like the ones in~(\ref{hami_int_qubit_rabi_type}) are taken into account. Using part of the second term in the right hand side of~(\ref{eq:mast:rabi:q}) we can calculate the following quantities:
\begin{subequations}
\begin{align}
{\rm tr}\big \{\mathcal{L}[f_1\sigma_z+f_2\sigma_x]\rho\,\sigma_x\big\}&=-2f_1^2\langle\sigma_x\rangle+2f_1f_2\langle\sigma_z\rangle,\\	
{\rm tr}\big \{\mathcal{L}[f_1\sigma_z+f_2\sigma_x]\rho\,\sigma_y\big\}&=-2(f_1^2+f_2^2)\langle\sigma_y\rangle,\\
{\rm tr}\big \{\mathcal{L}[f_1\sigma_z+f_2\sigma_x]\rho\,\sigma_z\big\}&=-2f_2^2\langle\sigma_z\rangle+2f_1f_2\langle\sigma_x\rangle.
\end{align}
\end{subequations}
We take these expressions to write the 
corresponding Bloch equations:
\begin{equation}\label{bloch_Rabi}
\frac{d\langle \vec{\sigma}\rangle}{dt}=\mathbb{B}\langle\vec{\sigma}\rangle,
\end{equation}
where $\vec{\sigma}=(\sigma_x,\sigma_y,\sigma_z)^{\intercal}$ and 
\begin{align}
\mathbb{B}=
\begin{pmatrix}
-\gamma_{\phi}&-\omega&\Omega\\
\omega&-(\gamma_{\phi}+\gamma)&0\\
\Omega&0&-\gamma
\end{pmatrix}.
\end{align}
Note that the following definitions have been used:
$\gamma_\phi= 2f_1^2\langle\{\sigma_-^B,\sigma_+^B\}\rangle$ 
$\gamma=2f_2^2\langle\{\sigma_-^B,\sigma_+^B\}\rangle$ and
$\Omega= 2f_1f_2\langle\{\sigma_-^B,\sigma_+^B\}\rangle$.
From~(\ref{bloch_Rabi}) we can interpret $\gamma_\phi$ as an effective dephasing rate, $\gamma$ as an effective decay rate 
and $\Omega$ can be seen as an effective pumping term. The equation \eqref{bloch_Rabi} is a homogeneous one, without any driving term inducing energy population or quantum coherence. 

\begin{figure}[t]
\begin{tikzpicture} 
  \node (img1)  {\includegraphics[width=.95\columnwidth]{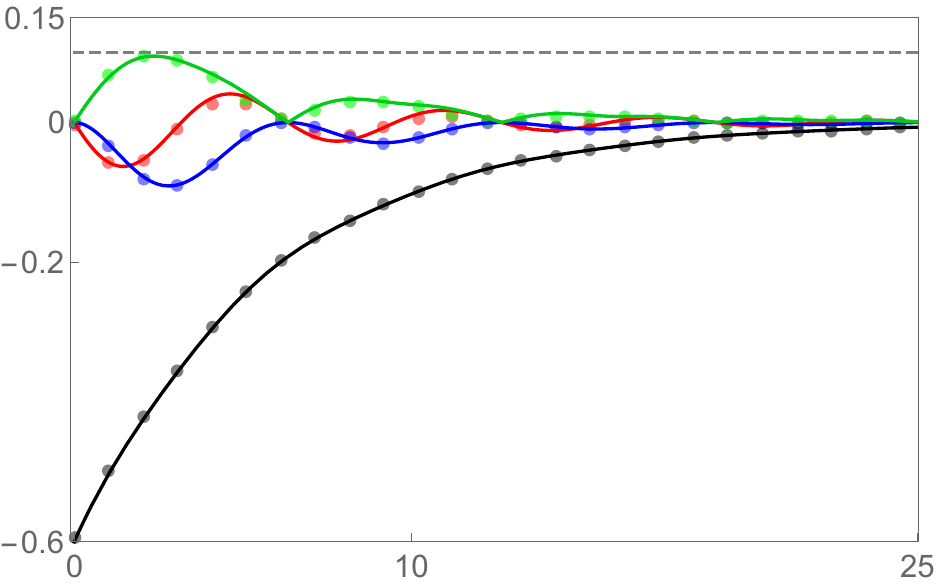}};
  \node[above=of img1, node distance=0cm, yshift=-6.3cm,xshift=.5cm] {$\omega t$};
  \node[above=of img1, node distance=0cm, yshift=-2.3cm,xshift=-2.8cm] {{\color{red}$\langle \sigma_x\rangle$}};
  \node[above=of img1, node distance=0cm, yshift=-3.2cm,xshift=-2.1cm] {{\color{blue}$\langle \sigma_y\rangle$}};
  \node[above=of img1, node distance=0cm, yshift=-4.3cm,xshift=-1.8cm] {{\color{black}$\langle \sigma_z\rangle$}};
  \node[above=of img1, node distance=0cm, yshift=-2.0cm,xshift=-1.1cm] {{\color{green}$\mathcal{C}$}};
  \node[above=of img1, node distance=0cm, yshift=-2.2cm,xshift=2.1cm] {{\color{black}$\overline{\mathcal{C}}_{\rm LHO}^{\rm{C-R}}$}};
  \end{tikzpicture}
\caption{Approximate evolution for the components
of the Bloch vector:
$\langle \sigma_x\rangle$ (red line),
$\langle \sigma_y\rangle$ (blue line) and
$\langle \sigma_y\rangle$ (black line).
The green line is the $l_1$-norm of coherence $\mathcal{C}$ and the black dashed line is the optimized maximum according to Eq.~\eqref{eq-trans-coh-approx-LHO}.
We have set $\sqrt{2}f_1=f_2$, $f_2=0.3$, $\omega=\omega_B=1$.
The initial state of the qubit probe is a mixed state 
such that $\langle\sigma_z(0)\rangle=-0.6$. 
Results from an exact numerical simulation of
the repeated (collision) interactions are shown 
as the tick opacity dots, where the time between each 
collision is set to $\tau=0.051$.}
\label{bloch_evolution}
\end{figure}
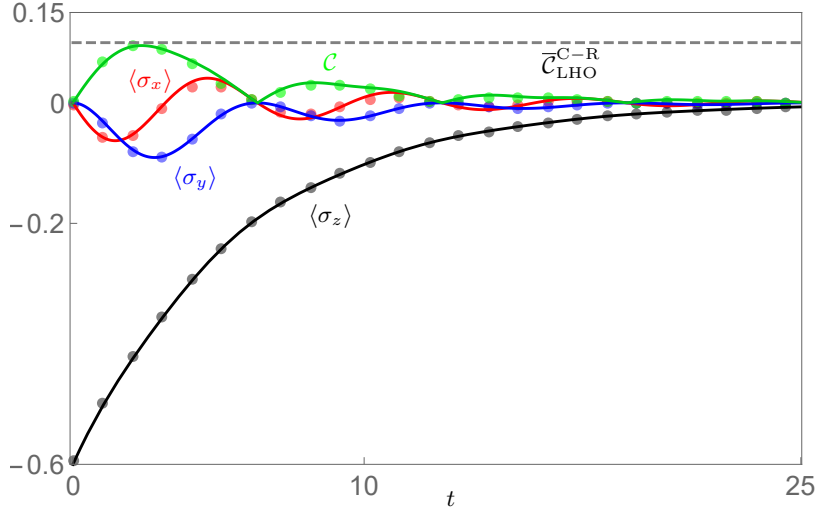

It is easy to check that the steady-state solution of
the Bloch vector is
$\langle\vec{\sigma}\rangle_{\rm ss}=(0,0,0)^{\intercal}$.
This means that the target qubit probe ends up into a mix 
state with equal probabilities. Therefore, no steady-state 
coherences can be generated in the qubit probe when
Rabi-type of interactions are considered as 
the orthogonal part of the system Hamiltonian.
However, during the time evolution or transient, it is still possible to show that a certain amount of coherences in the target qubit can be generated. To see this, using the Laplace transform method, we obtain the following approximated solutions for each component of the Bloch vector
\begin{eqnarray}
\langle\sigma_x(t)\rangle&\approx&\frac{z_0\Omega \exp(-3\gamma_\phi t/2)}{\sqrt{\omega^2-\Omega^2-\gamma_\phi^2/4}}\sin\Big(t\sqrt{\omega^2-\Omega^2-\gamma_\phi^2/4}\Big),
\nonumber\\ \label{approx_bloch_rabi_x} \\  
\langle\sigma_y(t)\rangle&\approx&\frac{z_0\Omega\omega\exp({-2\gamma_\phi t})}{\omega ^2-\Omega^2}\left[1-\cos\left(t\sqrt{\omega^2-\Omega^2}\right)\right],\label{approx_bloch_rabi_y} 
\nonumber\\ \\
\langle\sigma_z(t)\rangle&\approx&\frac{\Omega}{\omega}\langle\sigma_y(t)\rangle+z_0\exp(-2\gamma_\phi t),\label{approx_bloch_rabi_z}
\end{eqnarray}
where $z_0=\langle\sigma_z(0)\rangle$ and 
$\langle\sigma_x(0)\rangle=\langle\sigma_y(0)\rangle=0$ are the initial conditions of $\langle\vec{\sigma}\rangle$. 
The above expressions were obtained under the assumption 
$\gamma\approx 2\gamma_\phi$, corresponding to the choice of the values of coupling constants $f_2=\sqrt{2}f_1$. Additionally to this condition,
we have made the approximation $3\gamma_\phi\approx 2\gamma_\phi$, by assuming small coupling values $f_1$, $f_2$ with respect to $\omega$.
Therefore, (\ref{approx_bloch_rabi_x}-\ref{approx_bloch_rabi_z}) will be a good approximated solution of the Bloch vector if all these requirements are satisfied, see an example in 
Fig.~\ref{bloch_evolution}. These assumptions suggest that the more general form of the exponential arguments within these approximations is $\exp[-t(\gamma+\gamma_\phi)/2]$. To obtain results allowing for time optimized values of $\overline{C}_{\rm TS}$ and $\overline{\mathcal{P}}_{\rm TS}$, we neglect $\gamma_\phi$ and $\Omega$ with respect to $\omega$ in arguments of goniometric functions in Eqs.~\eqref{approx_bloch_rabi_x}, yielding 
\begin{eqnarray}
\langle\sigma_x(t)\rangle&\approx&\frac{z_0\Omega \exp[-t(\gamma+\gamma_\phi)/2]}{\omega}\sin(t\omega), \label{approx_bloch_rabi_x2} \\  
\langle\sigma_y(t)\rangle&\approx&\frac{z_0\Omega\exp({-t(\gamma+\gamma_\phi)/2})}{\omega}\left[1-\cos\left(t{\omega}\right)\right],\qquad\label{approx_bloch_rabi_y2}
\\
\langle\sigma_z(t)\rangle&\approx& z_0\exp[-t(\gamma+\gamma_\phi)/2].\label{approx_bloch_rabi_z2}
\end{eqnarray}
Such simplified time evolution allows for time-optimization of the coherence $\mathcal{C}\equiv |\langle\sigma_x(t)\rangle+i\langle\sigma_y(t)\rangle|$ and purity $\mathcal{P}\equiv (1+|\langle\vec{\sigma}\rangle|^2)/2$, yielding Eqs.~\eqref{eq-trans-coh-approx-TLS}-\eqref{eq-trans-coh-approx-LHO} and Eqs.~\eqref{eq-trans-purity-TLS}-\eqref{eq-trans-purity-LHO}. These results represent in fact weakly damped oscillations of the Bloch vector in the regime of small system-bath coupling constants $f_{1(2)}$.

The optimization procedure has the same ground in the case of RWA interaction, Eq.~\eqref{hami_int_qubit}, only the intermediate results are more cumbersome.

\section{Equations of motion for the target oscillator}\label{eqs_oscillaor_case}

Here we present the equations of motion of a target harmonic 
oscillator when this interacts, repeatedly, with a set 
of bath oscillators, as described in Sec.~\ref{SSC_oscillator}
of the main text. Once we have the equation of motion we will
obtain the expectation value of the oscillator's amplitude
$\langle a\rangle$ at the steady state.

First, it is easy to show that 
${\rm tr}\{\mathcal{L}[f_1 a^\dagger a +f_2 X_a]\rho a\}=-f_1^2\langle a\rangle/2-f_1f_2/2$,
and together with the master equation (\ref{mas_eq_osc}) 
of the main text, we obtain the following equation of motion
\begin{equation}\label{eq_moti_a_no_rwa}
\frac{d}{dt}\langle a\rangle=-i\omega_0\langle a\rangle-\frac{f_1^2}{2}\langle\{b,b^\dagger\}\rangle\langle a\rangle-\frac{f_1f_2}{2}\langle\{b,b^\dagger\}\rangle.
\end{equation}
Second, assuming the target oscillator is initially in an incoherent state,
then $\langle a(0)\rangle=0$ and one can find the time-dependent 
solution of the above equation given by 
$\langle a(t)\rangle=-(1-e^{-K_1t})K_2/K_1$, where
$K_1\equiv i\omega_0+f_1^2\langle\{b,b^\dagger\}\rangle/2$ 
and $K_2\equiv f_1f_2\langle\{b,b^\dagger\}\rangle/2$.
At the steady state the left hand side of Eq.~(\ref{eq_moti_a_no_rwa})
vanishes and $\langle a\rangle_{\rm ss}=-K_2/K_1$, substituting
$K_{1,2}$ in such ratio one obtains 
$\langle a \rangle_{\rm ss}=-f_1f_2/(f_1^2+i2\tilde{\omega})$
with $\tilde{\omega}\equiv \omega_0/\langle \{b,b^\dagger\}\rangle$. 

On the other hand, using the master equation (\ref{eq_maes_osc_rwa}) 
of the main text, it is also easy to show the following equation 
\begin{eqnarray}\label{a_eq:motion}
\frac{d}{dt}\langle a\rangle =-\Big(i\omega_0+\frac{f_1^2}{2}\langle\{b,b^\dagger\}\rangle+\frac{f_2^2}{2}\Big)\langle a\rangle
\nonumber\\
-\frac{f_1f_2}{2}\langle b^\dagger b\rangle-\frac{f_1f_2}{2}\langle a^2\rangle,
\end{eqnarray}
To solve the above equation first we need to find the expression for 
the equation of motion of $d\langle a^2\rangle/dt$, this will depend
on terms like $\langle a^3\rangle$ and so on. 
With this procedure we will end up with an infinite number of 
coupled linear differential equations. However, assuming weak 
coupling between the target oscillator and the bath elements, 
$f_{1,2}^2\ll \omega_0$, one may neglect the last term in the right hand 
side of (\ref{a_eq:motion}), which depends on powers $f_{1,2}^3$ 
and higher. Thus, an approximate steady solution for the 
target oscillator's amplitude is just
$\langle a\rangle_{\rm ss}={-f_1f_2\langle b^\dagger b\rangle}/{\big(i2\omega_0+{f_1^2}\langle\{b,b^\dagger\}\rangle+{f_2^2}\big)}$,
suggesting that SSC could be generated in the target 
oscillator only when the product $f_1\cdot f_2$ is nonzero.
Additionally, we require the bath oscillators being
in a thermal state with a temperature $T\neq 0$ such
that $\langle b^\dagger b\rangle\neq 0$.

%

%

\end{document}